\documentclass[final]{siamart0516}

\usepackage{lipsum}
\usepackage{amsfonts}
\usepackage{graphicx}
\usepackage{epstopdf}
\usepackage{algorithmic}
\usepackage{ amssymb }
\ifpdf
  \DeclareGraphicsExtensions{.eps,.pdf,.png,.jpg}
\else
  \DeclareGraphicsExtensions{.eps}
\fi

\newcommand{\TheTitle}{Self-similar solutions to isothermal shock problems} 
\newcommand{\TheAuthors}{S. C. Deschner, T. F. Illenseer, and W. J. Duschl}

\headers{\TheTitle}{\TheAuthors}

\title{{\TheTitle}\thanks{Accepted by SIAM Journal on Applied Mathematics, July 25, 2017.
}}

\author{
  Stephan C. Deschner\thanks{Institut f\"ur Theoretische Physik und Astrophysik (ITAP), Christian-Albrechts-Universit\"at zu Kiel
    (\email{sdeschner@astrophysik.uni-kiel.de}).}
  \and
  Tobias F. Illenseer\thanks{Institut f\"ur Theoretische Physik und Astrophysik (ITAP), Christian-Albrechts-Universit\"at zu Kiel (\email{tillense@astrophysik.uni-kiel.de}).}
  \and
  Wolfgang J. Duschl\thanks{Institut f\"ur Theoretische Physik und Astrophysik (ITAP), Christian-Albrechts-Universit\"at zu Kiel and Steward Observatory, The University of Arizona, 933 N. Cherry ave., Tucson, AZ 85721, USA (\email{wjd@astrophysik.uni-kiel.de}).}
}

\usepackage{amsopn}

\ifpdf
\hypersetup{
  pdftitle={\TheTitle},
  pdfauthor={\TheAuthors}
}
\fi

\begin{document}

\maketitle

\begin{abstract}
  We investigate exact solutions for isothermal shock problems in different one-dimensional geometries. These solutions are given as analytical expressions if possible, or are computed using standard numerical methods for solving ordinary differential equations. We test the numerical solutions against the analytical expressions to verify the correctness of all numerical algorithms.

  We use similarity methods to derive a system of ordinary differential equations (ODE) yielding exact solutions for power law density distributions as initial conditions. Further, the system of ODEs accounts for implosion problems (IP) as well as explosion problems (EP) by changing the initial or boundary conditions, respectively. 
  
  Taking genuinely isothermal approximations into account leads to additional insights of EPs in contrast to earlier models. We neglect a constant initial energy contribution but introduce a parameter to adjust the initial mass distribution of the system. Moreover, we show that due to this parameter a constant initial density is not allowed for isothermal EPs. Reasonable restrictions for this parameter are given.

  Both, the (genuinely) isothermal implosion as well as the explosion problem are solved for the first time.
\end{abstract}




\section{Introduction}
  
  Strong one-dimensional (1D) point-like explosions have been investigated thoroughly since the pioneering work of \cite{Sedov1946,Taylor1950}. Shock waves arising by reflection of a fluid on a solid wall were studied by Noh \cite{Noh1987}. Meyer \cite{Meyer1988} used Nohs solution for code verification whereby he also confirmed his findings with experimental data. Gehmeyr \cite{Gehmeyr1997} enhanced Nohs implosion problem (IP) by giving analytical solutions to $n$ shock reflections on rigid walls on both sides of a tube. Since decades, these solutions are used for testing CFD codes because resolving shock waves is a crucial feature for verifying conservation properties of inviscid, compressible fluids. Moreover, Sedovs explosion problem (EP) is still in use for simple modelling of Supernova Remnants (SNR) \cite{Allen2014,Leahy2016,Lerche1976,Solinger1975} in their early evolutionary states \cite{Vink2012,Woltjer1972}.

  Almost all previous works considered energy conservation with an ideal gas equation of state. To the authors knowledge, the IP was never considered to be isothermal. In case of the EP an unphysical diverging temperature gradient behind the shockwave \cite{Sedov1959, Solinger1975} occurs. Referring to this, Korobeinikov \cite{Korobeinikov1956,Korobeinikov1976} first considered isothermality with the assumption of a vanishing temperature gradient ahead of the shock wave. This concept leads to an adiabatic flow through the discontinuity and the principle of energy conservation therefore holds. Together with boundary conditions given by adiabatic Rankine-Hugoniot Conditions (RHC) \cite{Rankine1870, Hugoniot1887} well defined solutions exist. Nevertheless, above assumptions describe a mixed system of adiabaticity and isothermality and hence is not a genuinely isothermal system. Moreover, Lerche \cite{Lerche1977} stated that no valid isothermal self-similar blast-wave model provides applicability to SNR.

  In this work, we want to give solutions of both the genuinely isothermal IP and EP with just one governing set of equations. In the investigated cases, the fluid will cool down immediately on both sides of the discontinuity to ensure constant temperature throughout the system. This does violate the principle of energy conservation but avoids the additional assumption of a vanishing temperature gradient ahead of the shock. We use similarity methods and give a mathematical justification for the existence and appearance of the Lie invariants via a stretching transformation. We derive one set of equations governing both, the IP and the EP by simply changing the corresponding boundary conditions. 
  In case of the EP isothermal systems may lead to applicability to early phase SNR. Whilst many modern CFD codes are also able to model isothermal systems, corresponding test cases are merely rare. The results derived in this paper may expand the field of applicable test cases by a set of new solutions.

  We start the investigation by describing the model assumptions and the derivation of the Lie invariants together with required auxiliary conditions in \cref{sec:matcons}. The solutions are then discussed in \cref{sec:ssimsol} by investigating the phase diagrams with respect to different geometries and free parameters to the system. The results are summarized in \cref{sec:summary} and the \cref{anasolIP,anasolEP} give analytical solutions to the IP and EP in planar geometry.

  \begin{figure}
   \centerline{\includegraphics{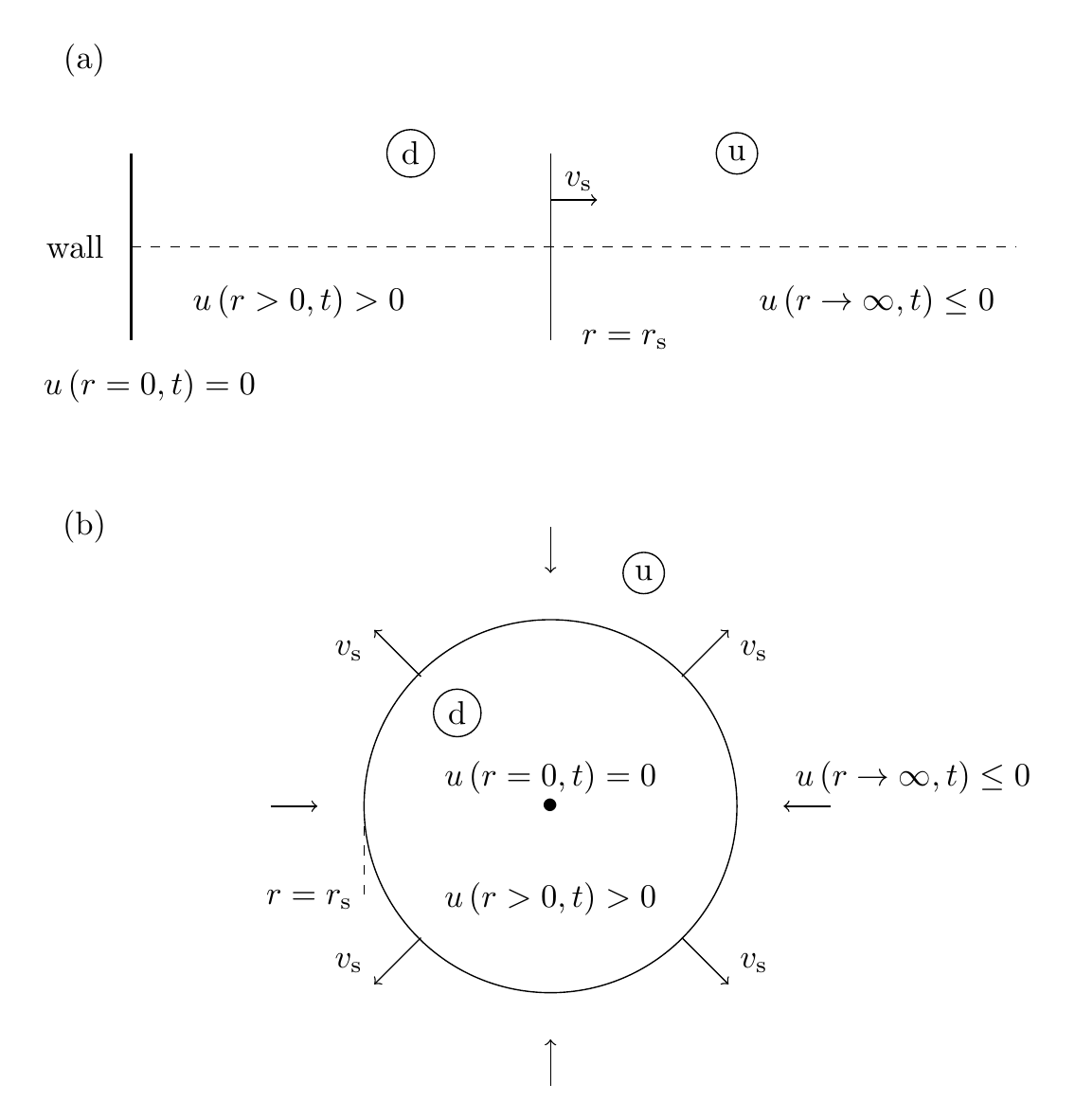}}
      \caption{Simplest pictures of 1D shocks in different coordinates. Figure (a) shows a shock front in plane geometry for a shock arising by fluid reflection at the wall (IP). Figure (b) shows the same in polar and spherical coordinates with reflecting boundary conditions at the center. The down ($\mathrm{d}$)- and upstream ($\mathrm{u}$) regions are labeled with respect to the shock front and change to $\left(d, u\right)^\mathrm{IP}\mapsto\left(u, d\right)^\mathrm{EP}$ while considering explosion problems. In this case, the explosion propagates into a quiescent region with $u\left(r\rightarrow\infty, t\right)=0$.
              }
         \label{fig:Figtube}
   \end{figure}
%

\section{Mathematical considerations}\label{sec:matcons}
  \Cref{fig:Figtube}a shows a sketch of a 1D shock tube filled with an inviscid, compressible isothermal fluid. We split the spatial domain into three distinct regions, an upstream region $\left( \mathrm{u}\right)$, a downstream region $\left( \mathrm{d}\right)$ and an infinitesimally thin region around the shock at $r=r_{\mathrm s}$. The sketch further shows a rigid wall on the left hand side where the inflowing gas with velocity $u\left(r, t\right)<0$ gets reflected. Hence, a shock wave arises propagating to the right with shock speed $v_{\mathrm s}$. Now, the velocity in the downstream region becomes $u\left(r, t\right)\geq0$, but always equals zero at the origin where the fluid necessarily stagnates. The shock itself becomes apparent through a discontinuity or jump. 
  
  To draw a more general picture we can easily enlarge the above scenario by introducing 1D polar coordinates or even 1D spherical coordinates instead of 1D cartesian coordinates \cite{Gehmeyr1997} with respect to the corresponding symmetries (see \cref{fig:Figtube}b). The basic equations are very similar and can be  distinguished by the exponent $\alpha\in\left\{0,1,2\right\}$ of the radius (see \Cref{eq:origsys_c,eq:origsys_i}). The rigid wall changes to the center of the disk or sphere and the fluid flows radially.

  In the case of the EP, the only difference lies in the change of the domains $\left(\mathrm{u}, \mathrm{d}\right)\mapsto\left(\mathrm{d}, \mathrm{u}\right)$ and the shock propagates into the quiescent downstream region $u\left(r\rightarrow\infty, t\right)=0$.

  Mathematically, the up- and downstream regions are described with the 1D continuity and Euler equations. The jump itself must be determined via the Rankine-Hugoniot Conditions (see \cref{secRHC}). We further use the isothermal condition $\left(T=\mathrm{const}\right)$ and assume an ideal gas equation of state. Hence, $P/\rho=\mathrm{const.}$, or
  \begin{equation}
   P=\rho c_{\mathrm s}^2\label{eq:isocond}
  \end{equation}
  where $P$ is the pressure, $\rho$ the density and $c_\mathrm{s}$ the constant speed of sound, leading to a violation of the principle of energy conservation.
 
  The generalized equations e.g. \cite{Landau1974eng} together with equation \cref{eq:isocond}, the velocity normalized to the speed of sound and the density normalized to some characteristic density of the flow $\hat \rho$ can be written as a system of two nonlinear partial differential equations (PDE)
  \begin{subequations}
  \begin{align}
   \partial_\tau\rho+\frac{1}{r^\alpha}\partial_r\left(r^\alpha\rho u\right)&=0\label{eq:origsys_c}\\
   \partial_\tau u+u\partial_r u +\frac{1}{\rho}\partial_r\rho &=0.\label{eq:origsys_i}
  \end{align}
  \end{subequations}
  The density $\rho=\rho\left(r,\tau\right)$ and the radial velocity $u=u\left(r,\tau\right)$ are functions of the spatial coordinate $r$ and the normalized time $\tau=c_\mathrm{s}t$.

\subsection{The reduced differential equations}\label{redmeth}
  Self-similar solutions to a set of partial differential equations (PDEs) must satisfy two constraints. First, a group of transformations exists reducing the set of PDEs to a set of ordinary differential equations (ODEs) by linking the independent variables $\left(r, \tau\right)$ to one new independent variable $\xi\left(r, \tau\right)$. Second, the reduced set of ODEs must also yield a well-posed problem. This is achieved if the auxiliary conditions reduce to a well-posed boundary value problem, solely expressible in terms of the similarity variables \cite{Ames1965,Bluman2002}. 

  In general there are several ways to find transformation groups, reducing a PDE to an ODE for example with a dimensional analysis \cite{Sedov1946,Sedov1959} or with the method of separation of variables \cite{Ames1965}. However, these methods may not provide self-similarity necessarily, and are based on some ``intuitive'' guess to determine the transformation group \cite{Ames1965,Sedov1959}.

  In this work, we want to apply the general method of stretching transformations ST (or Birkhoff's first method) \cite{Ames1965,Logan2008}. The ST applied to (systems of) PDEs yields self-similar solutions if there exists a group of point transformations leaving the set of PDEs invariant. This is meant as a duality principle: the method only provides a transformation- or stretching group (SG) if similarity exists and the other way around. Hence, the striking features of the ST are its simplicity and clarity in determining invariance and the corresponding SG at once. This method also excludes this ``intuitive'' guess used in alternative methods \cite{Ames1965}.

  Formally, applying the stretching transformation to \Cref{eq:origsys_c,eq:origsys_i} means to check under which conditions the point transformations
  \begin{equation}
   \bar \tau=\lambda^{a}\tau,~~~\bar r=\lambda^{b}r,~~~\bar \rho=\lambda^{c}\rho,~~~\bar u=\lambda^{d}u,\label{eq:stretch}
  \end{equation}
  leave the system of PDEs invariant. Here, $a,b,c,d$ are constants and $\lambda \in \mathbb{R}$ is the group parameter. Invariance is found for 
  \begin{equation}
   a=b\nonumber,~c\neq0,~d=0.\nonumber
  \end{equation}
  Hence, the solutions are arbitrary functions of the group invariants given by the similarity variable  
  \begin{equation}
   \xi\left(r,\tau\right)=\frac{r}{\tau},\label{eq:simvar}
  \end{equation}
  the condition for the density
  \begin{equation}
   \rho\left(r, \tau\right)=S\left(\xi\left(r, \tau\right)\right)r^{-\kappa}\label{eq:densdist}
  \end{equation}
  with $\kappa\equiv b/c$ together with the condition for the velocity
  \begin{equation}
   u\left(r, \tau\right)=u\left(\xi\left(r, \tau\right)\right). \label{eq:veltrans}
  \end{equation}
  The above group invariant solutions (\Cref{eq:simvar,eq:densdist,eq:veltrans}) inserted into the PDEs (\cref{eq:origsys_c,eq:origsys_i}) yields the basic system of ODEs\footnote{We use the differential transformations $\partial_{\tau}=-\frac{\xi}{\tau}\frac{\mathrm{d}}{\mathrm{d}\xi}$, $\partial_{r}=\frac{\xi}{r}\frac{\mathrm{d}}{\mathrm{d}\xi}$}
  \begin{subequations}
  \begin{align}
   \frac{\mathrm{d}u}{\mathrm{d}\xi}&=\frac{u\alpha-\kappa\xi}{\xi\left(\left(\xi-u\right)^2-1\right)},\label{eq:odev}\\
   \frac{\mathrm{d}S}{\mathrm{d}\xi}&=\frac{Su\left(\alpha-\kappa\right)\left(\xi-u\right)-S\kappa}{\xi\left(\left(\xi-u\right)^2-1\right)}.\label{eq:odes}
  \end{align}
  \end{subequations}
  Alternatives to the group invariants emerge by a simple scaling with an arbitrary constant, but results in the same system \cref{eq:odev,eq:odes} \cite{Ames1965}. Solutions to the PDE are mapped from solutions of the ODE with the invariants of the SG \cref{eq:simvar,eq:densdist,eq:veltrans}. Moreover, invariance reduces the independent variable $\tau\in \mathbb{R}_0^{+}$ to a simple constant. The concerning auxiliary conditions are derived in \cref{initbnd}. Realizing that the velocity equation \cref{eq:odev} is decoupled from the function $S$ allows to investigate the behavior of the solution graphically (see \cref{sec:ssimsol}).



  \subsection{Auxiliary conditions}\label{initbnd}
 The constraint of a well-posed boundary value problem requires that the initial (IC) and boundary conditions (BC) of the PDEs coincide \cite{Ames1965}. The PDE \cref{eq:origsys_c,eq:origsys_i} requires one IC ($f\left(x, t=0\right)=f_\mathrm{i}$) and two BCs ($f\left(x_{0\left(\infty\right)}\right)=f_{0\left(\infty\right)}$) for each $f\in\left\{u,\rho\right\}$ yielding six auxiliary conditions. The limits of the similarity variable \cref{eq:simvar}
  \begin{align}
    \xi\left(r,\tau\right)\rightarrow \left\{
    \begin{array}{ll}
      0,  & \mbox{if\ }\quad r\rightarrow 0~\mbox{or\ } \tau\rightarrow\infty\\[2pt]
      \infty, & \mbox{if\ }\quad r\rightarrow\infty~\mbox{or\ } \tau\rightarrow0\label{eq:iccases}
    \end{array}\right.
  \end{align}
  combine the spatial conditions at infinity with the conditions for $\tau=0$, or more specifically, the IC with the outer BC
  \begin{align}
   u_\mathrm{i}\left(r\right)=\lim_{\tau\to0}u\left(r,\tau\right)&=\lim_{\xi\to\infty}u\left(\xi\right)=u_\infty\\
   \rho_\mathrm{i}\left(r\right)=\lim_{\tau\to0}\rho\left(r,\tau\right)&=r^{-\kappa}\lim_{\xi\to\infty}S\left(\xi\right)=r^{-\kappa} S_\infty.\label{eq:IC_out}
  \end{align}
  A constant initial velocity field for the dynamical system \cref{eq:origsys_c,eq:origsys_i} only exists in the limit $\lim_{\tau\to0}u\left(r,\tau\right)\rightarrow u_\mathrm{i}$. This limit may be explained by e.g.  \cref{fig:figure7} showing the time evolution of the velocity in the $u\left(r, \tau\right)$-plane. The shorter the time, the smaller the osculating circle to the function describing the upstream region until it is ending up in a $\delta$-distribution-like initial condition. The initial density distribution equation \cref{eq:densdist} is given by a power law depending on radius and the constant $S_{\infty}$. We may argue, that the physically relevant profiles are due to only non-negative values of $\kappa$, because we want the majority of the fluid located at the inner regions. Hence, we restrict the value $\kappa\in \mathbb{R}_0^{+}$ and refer the reader to the end of \cref{anasolIP} and \cref{secgenrestrict} for additional mathematical limitations required for the solvability of the system. The constant $S_{\infty}>0$ is a free parameter. Without loss of generality we set\footnote{We want to remind that \Cref{eq:odev,eq:odes} are dimensionless equations}
  \begin{equation}
    S_{\infty}=1\label{eq:sbnd}
  \end{equation}
  for both the implosion as well as the explosion problem. This BC is the only condition for fixing the density solution because we either need to know the inner BC ($S_0$) or the outer BC ($S_\infty$) together with the jump position $\xi_0$. The jump position is given by the RHC which we will discuss in \cref{secRHC}.

  We derive the inner BC for the velocity distribution (\Cref{eq:veltrans}) by a linearization of \Cref{eq:odev}
  \begin{equation}
    \frac{\mathrm{d}u}{\mathrm{d}\xi}\approx\frac{\kappa\xi-u\alpha}{\xi} \mbox{,\ \ for\ }\quad 0<\xi\ll1\quad\mbox{and}\quad 0<\left|u\right|\ll1\label{eq:appr_u}
  \end{equation}
  and find the approximate solution
  \begin{equation}
   u\left(\xi^{\ll1}\right)=u_0=\frac{\kappa\xi}{\alpha+1}+c_{0}\xi^{-\alpha}.\label{eq:solv_ibnd}
  \end{equation}
  The constant $c_0$ has to be zero, because the fluid must halt at the wall or the center (see \cref{fig:Figtube}a, b). To satisfy \Cref{eq:solv_ibnd} we need the inner starting position $\xi\neq0$ and set $\left.\xi\right|_0=10^{-3}$, being valid for both, the IP and EP. 
  
  The outer velocity BCs are fixed in the same manner, whereas we define the dependent variable $\eta\doteq\xi^{-1}$ such that $\eta\rightarrow 0 \mbox{\ for\ } \xi\rightarrow \infty$. This enables us to give well defined and finite numbers for the BCs. The velocity equation \cref{eq:odev} then becomes
  \begin{equation}
  	\frac{\mathrm{d}u}{\mathrm{d}\eta}=\frac{\kappa-u\alpha\eta}{\left(1-u\eta\right)^2-\eta^2}\label{eq:veleta}
  \end{equation}
  and the solution of the approximate equation
  \begin{equation}
  	\frac{\mathrm{d}u}{\mathrm{d}\eta}\approx \kappa-u\alpha\eta\mbox{,\ \ for\ }\quad 0<\eta\ll1\quad\mbox{and}\quad -1\leq\left|u\right|\leq0
  \end{equation}
  is written in terms of the imaginary error function $\mathrm{erfi\left(z\right)}$
  \begin{equation}
  	u\left(\eta\right)=e^{\frac{-\alpha\eta^2}{2}}\left(c_1+\sqrt{\frac{\pi}{2\alpha}}\kappa\mathrm{\ erfi}\left(\sqrt{\frac{\alpha}{2}}\eta\right)\right)
  \end{equation}  
  with the constant of integration $c_1$. The IP requires a constant negative inflow velocity complying with \cite{Noh1987}, and the EP a vanishing velocity field as $\eta=\xi^{-1}\rightarrow0$ complying with \cite{Sedov1946}. Hence, the BCs are fixed with
  \begin{align}
    u_{\infty}&= \lim_{\eta\to0}u\left(\eta\right)\equiv c_1=\left\{
    \begin{array}{ll}
      -1 & \mbox{for\ IP}  \\
       0 & \mbox{for\ EP}
    \end{array}\right. .\label{eq:ubnd}
  \end{align}
  Applicability to IPs or EPs requires the RHC to be satisfiable under above BCs. Therefore, the outer BCs do not yield any loss of generality. The formerly six auxiliary conditions have been reduced to the two conditions on $\left(u_\infty,S_\infty\right)$ stating a well posed boundary value problem to the set of ODEs \cref{eq:odev,eq:odes}. The conditions $\left(u_0,S_0\right)$ emerge from solutions of the ODEs and application of the RHC.
  
  
  \subsection{Isothermal Rankine-Hugoniot Conditions}\label{secRHC}
  As indicated in \Cref{sec:matcons}, the RHC are required to describe the discontinuity created by the shockwave. These conditions are only valid in an infinitesimal area at the discontinuity and connect the adjacent states of the up- and downstream regions. To derive the jump conditions we refer the reader to the books of \cite{Courant1972} and with respect to isothermality to \cite{Clarke2003}. The RHC rewritten with the relations\footnote{We wish to remind the reader that equations (\cref{eq:origsys_c}, \cref{eq:origsys_i}) are normalized to the speed of sound $c_{\mathrm{s}}$ in contrast to \cite{Korobeinikov1956}} \cref{eq:simvar,eq:densdist} are given by
  \begin{eqnarray}
   S_\mathrm{u}\mathcal{U}_\mathrm{u}&=&S_\mathrm{d}\mathcal{U}_\mathrm{d}\nonumber\\
   S_\mathrm{u}\left(\mathcal{U}_\mathrm{u}^2+1\right)&=&S_\mathrm{d}\left(\mathcal{U}_\mathrm{d}^2+1\right),\nonumber
  \end{eqnarray}
  where
  \begin{equation}
   \mathcal{U}_\mathrm{u,d}=u_\mathrm{u,d}-\xi_0\label{eq:comov}
  \end{equation}
  denote the velocities in the comoving frame with respect to the discontinuity. The velocities $u_\mathrm{u,d}$ are the corresponding velocities in the inertial frame of reference and $\xi_0=\left.\xi\right|_{\left\{r_\mathrm{s},\tau_\mathrm{s}\right\}}$ yields the shock position. These conditions are independent of the radial distance because they vanish in the limit of an infinitely narrow strip and hence are also independent of the number $\alpha$ which distinguishes between the different coordinates. The RHC can be solved for the upstream states 
  \begin{equation}
   \mathcal{U}_\mathrm{u}=\mathcal{U}_\mathrm{d}^{-1},~~S_\mathrm{u}=S_\mathrm{d}\mathcal{U}_\mathrm{d}^2,\nonumber
  \end{equation}
  if $S_\mathrm{u}\neq S_\mathrm{d}$ and $S_\mathrm{u}, S_\mathrm{d}>0$. These conditions are symmetric, hence exchanging the indices does not change the equations (see \cref{secBW}). With the help of \Cref{eq:comov} one obtains the RHCs with respect to the inertial frame (see also \cref{fig:Figtube})
  \begin{subequations}
  \begin{align}
    u_{\mathrm{u}}&=\frac{1+u_{\mathrm{d}}\xi_0-\xi_0^2}{u_{\mathrm{d}}-\xi_0}\label{eq:RHCV}\\
    S_{\mathrm{u}}&=S_{\mathrm{d}}\left(u_{\mathrm{d}}-\xi_0\right)^2.\label{eq:RHCS}
  \end{align}
  \end{subequations}
  We emphasize that the numbers $u_{\mathrm{u,d}}, S_{\mathrm{u,d}}$ must satisfy the solutions of \Cref{eq:odev,eq:odes}. To compute values for the RHC we use \texttt{Octave}\footnote{A high-level interpreted language for numerical calculations under the GNU General Public License \cite{Eaton2014}.} with a 4\textsuperscript{th} order Adams-Moulton predictor-corrector scheme \cite{Hairer1993} and start solving the ODE \cref{eq:odev} at the inner boundary \cref{eq:solv_ibnd} up to some test value $\tilde\xi_0=\tilde\eta_0^{-1}$ as an initial guess for the jump position. This gives the number $u_\mathrm{d}(\tilde\xi_0)=\tilde u_\mathrm{d}$. Then we integrate Eq. \cref{eq:veleta} from the outer BC \cref{eq:ubnd} up to the value $u_\mathrm{u}(\tilde\eta_0)=\tilde u_\mathrm{u}$ and application of the RHC \cref{eq:RHCV} yields $\tilde u_\mathrm{d}^*$. We are now able to compute the relative error
  \begin{equation}
    \epsilon=\left|\frac{\tilde u_\mathrm{d} - \tilde u_\mathrm{d}^*}{\tilde u_\mathrm{d}^*}\right|\leq10^{-4}\label{eq:errbnd}
  \end{equation}
  with $\epsilon$ as a sufficient error bound. By checking over- or under estimation of the value $\tilde\xi_0$ we increase or decrease the next guess by $10\%$, staying close to the relevant region\footnote{The initial guess for $\tilde\xi_0$ is chosen to be around the singular point $\mathcal{D}_2$, which also is the domain of interest.} (see e.g. \cref{polshperea1k0}). The procedure then is repeated until a sufficient $\epsilon$ is found and hence converges to the exact RHC. The numbers for some particular solutions computed for different geometries are given in \cref{tab:numerRHC}.
  \begin{table}
  \caption{Numerical values for the RHC corresponding to the IP and different numbers for $\kappa$.
}
        \label{tab:numerRHC}
\centering
    \def~{\hphantom{0}}
     \begin{tabular}{|c|c|c|c|c|c|c|}\hline
       $\kappa$ & $\alpha$ & $\xi_0$ & $u_{\mathrm{u}}$ & $u_{\mathrm{d}}$ & $S_{\mathrm{u}}$ & $S_{\mathrm{d}}$\\ \hline
       0 & 0 & 0.6181 & -1.0 & 0.0 & 1.0 & 2.61780 \\
         & 1 & 0.7371 & -0.6196 & 0.0 & 2.6802 & 4.9328 \\
         & 2 & 0.8376 & -0.3562 & 0.0 & 5.4619 & 7.7844 \\ \hline
     1/3 & 0 & 0.9388 & -0.7906 & 0.3606 & 0.8507 & 2.5447 \\
     1/2 &   & 1.1905 & -0.7262 & 0.6688 & 0.8175 & 3.0030 \\
     2/3 &   & 1.5829 & -0.7017 & 1.1452 & 0.8004 & 4.1777 \\
      \hline
       1 & 1 & 1.3657 & -0.3028 & 0.7663 & 1.2373 & 3.4443 \\
         & 2 & 1.2594 & -0.0445 & 0.4925 & 2.1326 & 3.6260\\ \hline
     \end{tabular}
 \end{table}
\newline
We derive analytical solutions to both, the IP and the EP in slab geometry in \Cref{anasolIP,anasolEP}. The numerical values from \cref{tab:numerRHC} are checked against the analytical values derived by the corresponding methods with 
\begin{equation}
 \delta X=\frac{X_\mathrm{num}-X_\mathrm{analyt}}{X_\mathrm{analyt}}
\end{equation}
for every $X\in\left\{\xi_0,u_\mathrm{u},u_\mathrm{d},S_\mathrm{u},S_\mathrm{d}\right\}$. The results are summarized in \Cref{tab:vergleichRHC} showing a deviation of about $\delta X\approx 10^{-5}\mbox{\ to\ }10^{-6}$ which nicely demonstrates the accuracy of the described numerical method. The relative error of the numerical integration method is set to $10^{-6}$, such that the error bound \eqref{eq:errbnd} is well suited. The error bound is primarily justified by the precision of $\delta X$ but also by the computing time of the algorithm. 


\section{Self-similar solutions}\label{sec:ssimsol}
  We discuss the general method for solving \Cref{eq:odev,eq:odes} numerically. We give the solutions for $\kappa=0$ and $\kappa=1$ as examples in the different coordinates ($\alpha\in\left\{0,1,2\right\}$) and discuss further restrictions to $\kappa$. Solutions for the genuinely isothermal explosion problem in spherical coordinates are considered afterwards.

\subsection{Numerical solutions in arbitrary coordinates}\label{kappa=0}
  We achieve numerical solutions for all possible cases by numerical integration of the ODEs \cref{eq:odev,eq:odes}. We take advantage of the library \texttt{ode}\footnote{Function library for numerical integration of systems of ODEs under the GNU General Public License} \cite{Tufillaro1992}, which enables us to compute solutions with well known numerical solvers and with arbitrary accuracy. We chose the 4\textsuperscript{th} order Adams-Moulton scheme again (see \cref{secRHC}) and start the integration at the discontinuity $\mathcal{U}_\mathrm{u,d}\left(\xi=\xi_0\right)$ up to the origin $\xi=0$ and to $\left.\xi\right|_\infty=1000$. Again we do not lose uniqueness because we use the BCs to compute the RHCs. In the following, all solutions are plotted in the physical variables $\rho\left(r,\tau\right),~u\left(r,\tau\right)$, whereas we solve the dimensionless equations (the speed of sound is set to $c_{\mathrm{s}}=1$).
  
\begin{figure}
      \centerline{\includegraphics{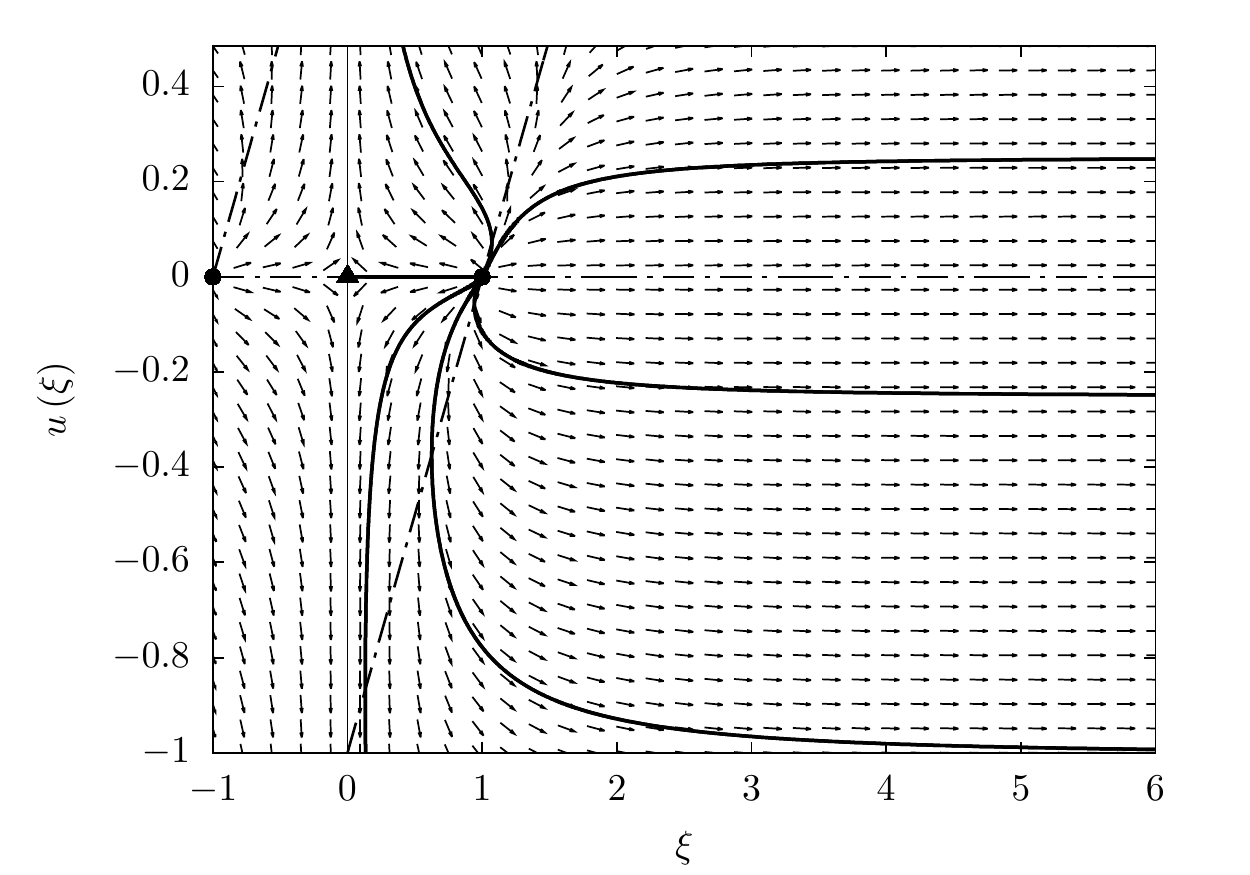}}
      \caption{Phase diagram for polar coordinates and constant density distribution ($\alpha=1, \kappa=0$). The thin solid curve denotes the separation of the domain by $\xi=0$. The dashed dotted lines are singular lines. Their intersection gives a saddle (triangle) and an unstable improper node (dots). Thick lines starting at singular points are selected solutions.
	      }
      \label{fig:pp_a1k0}
    \end{figure}

  To investigate the behavior of the solutions to subject certain values for $\kappa$ and BCs we emphasize some general properties by analyzing the phase diagrams. The nominator and denominator function (\Cref{eq:odev}) gives expressions for critical lines and critical points. We find a infinite or zero slope of the velocity field along the singular lines
    \begin{equation}
	    u\left(\xi\right)=\frac{\kappa\xi}{\alpha},~~u\left(\xi\right)=\xi\pm1,~~
	    \xi=0,\label{eq:negsign}
    \end{equation}
    where the negative sign in \cref{eq:negsign} accounts for intersections in the positive half-plane as long as $\kappa\leq\alpha$. For $\kappa>\alpha$ the intersection is due to the function with the positive sign. The last condition in \cref{eq:negsign} leads to diverging behavior of the solutions and therefore a nullcline. The positions of critical points are determined as the intersection of the singular lines at
    \begin{equation}
	    \mathcal{D}_1=\left(0,0\right),~~~\mathcal{D}_2=\left(\frac{\pm\alpha}{\alpha-\kappa},\frac{\pm\kappa}{\alpha-\kappa}\right).\label{eq:singpnt}
    \end{equation}
    The second point depends on geometry and on the exponent of the initial density distribution. By a linearization of \Cref{eq:odev} we rewrite the ODE in dependence of some parameter $t$ to
    \begin{equation}
    \frac{\mathrm{d}}{\mathrm{d}t}	
    \begin{pmatrix}
	    \xi  \\ u
    \end{pmatrix}=
    \begin{pmatrix}
	    u^2-4u\xi+3\xi^2-1 & 2u\xi-2\xi^2\\
	    -\kappa 				  & \alpha
    \end{pmatrix}_{\mathcal{D}_{1,2}}
    \begin{pmatrix}
	    \xi \\ u
    \end{pmatrix}
    =\mathbf{J}_{\mathcal{D}_{1,2}}^{\left(\alpha,\kappa\right)}
    \begin{pmatrix}
	    \xi \\ u
    \end{pmatrix}\label{eq:spanalysis}
    \end{equation}
    where $\mathbf{J}_{\mathcal{D}_{1,2}}^{\left(\alpha,\kappa\right)}$ is the Jacobian matrix at the positions $\mathcal{D}_{1,2}$. An analysis of the eigenvalues and eigenvectors \cite{Logan2008} of $\mathbf{J}_{\mathcal{D}_{1,2}}^{\left(\alpha,\kappa\right)}$ at the related positions enables us to characterize the critical points and hence the behavior of the solutions next to them. The eigenvectors are denoted with $\boldsymbol{x}_{1,2}$ for $\mathcal{D}_1$ and $\boldsymbol{z}_{1,2}$ for $\mathcal{D}_2$. The eigenvector $\boldsymbol{x}_{1}=\left(0,1\right)^{\mathrm{T}}$ is similar in all cases and describes the unstable diverging behavior at $\mathcal{D}_1$ yielding a saddle point. A solution of \Cref{eq:spanalysis} is given by \Cref{eq:solv_ibnd} as inner BC, so we can characterize the behavior of the velocity at the origin.
    \begin{table}
    \caption{Comparison between the numerical (\cref{secRHC}) and analytical (\cref{anasolIP,anasolEP}) computation of the RHC. We state very good agreement and hence correctness of the numerical method.
}
	  \label{tab:vergleichRHC} 
    \centering
      \def~{\hphantom{0}}
      \begin{tabular}{|c|c|c|c|c|c|c|c|}\hline
	  & $\kappa$ & $\alpha$ & $\delta\xi_0$ & $\delta u_{\mathrm{u}}$ & $\delta u_{\mathrm{d}}$ & $\delta S_{\mathrm{u}}$ & $\delta S_{\mathrm{d}}$\\ \hline
	& 1/3 & 0 & $3.29 \cdot 10^ {-5}$ & $2.01 \cdot 10^ {-6}$ & $4.28 \cdot 10^ {-5}$ & $3.59 \cdot 10^ {-6}$ & $5.72 \cdot 10^ {-5}$ \\ 
	IP & 1/2 &   & $2.63 \cdot 10^ {-5}$ & $3.44 \cdot 10^ {-6}$ & $3.15 \cdot 10^ {-5}$ & $4.23 \cdot 10^ {-6}$ & $4.37 \cdot 10^ {-5}$ \\ 
	& 2/3 &   & $2.19 \cdot 10^ {-5}$ & $4.18 \cdot 10^ {-6}$ & $2.38 \cdot 10^ {-5}$ & $3.68 \cdot 10^ {-6}$ & $3.78 \cdot 10^ {-5}$ \\ \hline 
	& 1/3 & 0 & $1.55 \cdot 10^ {-5}$ & $1.37 \cdot 10^ {-5}$ & $2.27 \cdot 10^ {-5}$ & $5.95 \cdot 10^ {-6}$ & $1.03 \cdot 10^ {-5}$ \\ 
	EP & 1/2 &   & $1.62 \cdot 10^ {-5}$ & $3.09 \cdot 10^ {-6}$ & $2.45 \cdot 10^ {-5}$ & $3.56 \cdot 10^ {-6}$ & $1.36 \cdot 10^ {-6}$ \\ 
	& 2/3 &   & $1.92 \cdot 10^ {-5}$ & $7.16 \cdot 10^ {-6}$ & $2.67 \cdot 10^ {-5}$ & $5.92 \cdot 10^ {-7}$ & $1.24 \cdot 10^ {-5}$  \\ \hline 
      \end{tabular}
    \end{table}
    In general, the singular lines divide the phase space in disjoint regions. None of the solutions from one region connects continuously to a solution in the other region \cite{Sedov1959}. Thus a solution stretching across the whole spatial domain must contain a discontinuity. Possible solutions shown in the phase diagrams \cref{fig:pp_a1k0,fig:pp_a2k0,fig:pp_a1k1,fig:pp_a2k1,fig:pp_blast_k0,fig:pp_kink} are no continuous solutions, they simply end at some joint position at the singular line.
 
   \subsubsection{Cartesian coordinates and \texorpdfstring{$\kappa=0$}{}}
    This constitutes the simplest case because the ODEs \cref{eq:odev,eq:odes} reduce to constant solutions separated only by the discontinuity. Hence, we are left with calculating the RHCs \cref{eq:RHCV,eq:RHCS} using the method described in \cref{secRHC}. All the numbers are collected in \cref{tab:numerRHC}. The solutions, yielding a simple step function are plotted in figure \cref{fig:solk0comb}a,b. The graphs show them back transformed with the similarity variable $\xi\mapsto r=\xi\cdot\tau$ (\Cref{eq:simvar}), so the shock position refers to $r=r_\mathrm{s}$ at different times $\tau$ again. For comparison, we mention the solutions for non-isothermal cartesian shocktubes \cite{Meyer1988, Gehmeyr1997}.

   \subsubsection{Polar and spherical coordinates and \texorpdfstring{$\kappa=0$}{}}\label{polshperea1k0}
\begin{figure}
  \centerline{\includegraphics{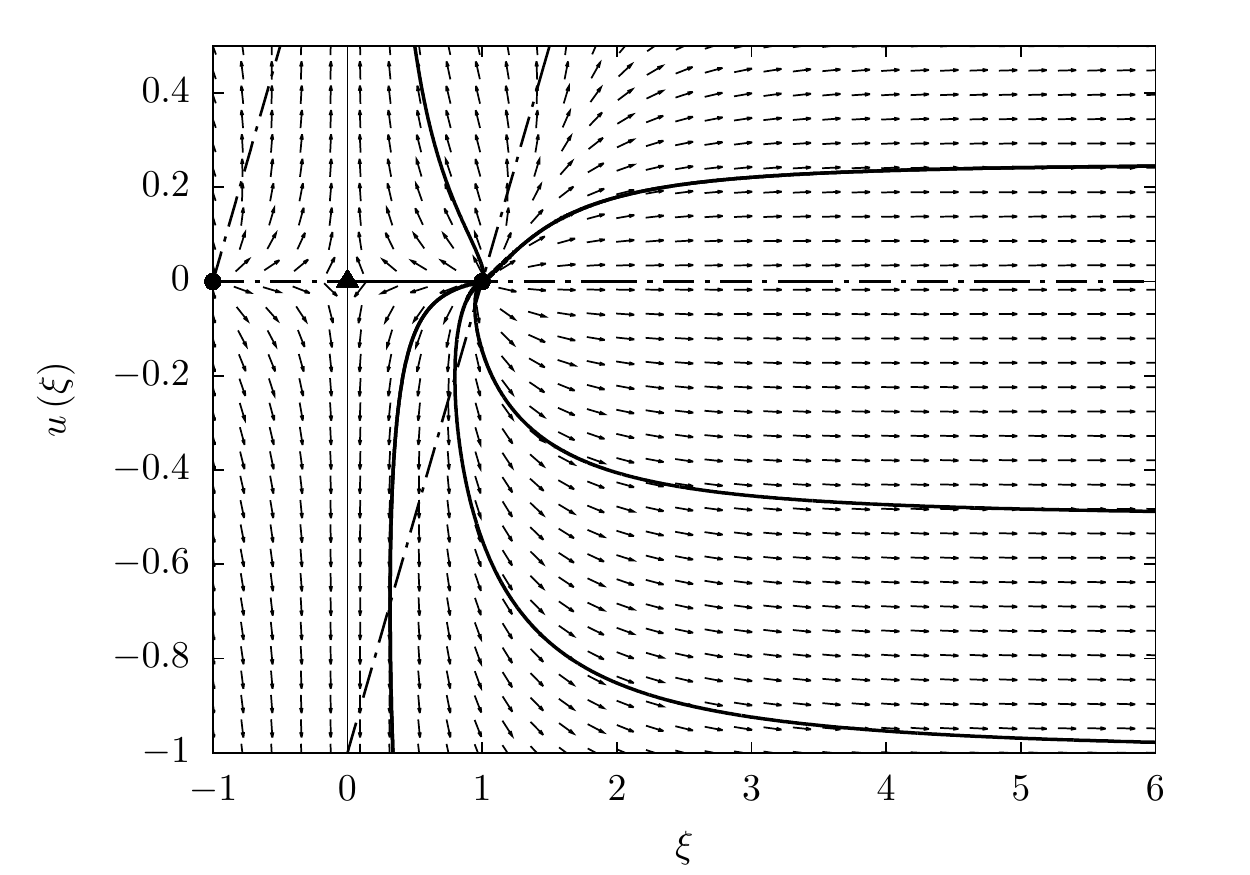}}
      \caption{Phase diagram for spherical coordinates ($\alpha=2, \kappa=0$). The singular lines and points (a saddle (triangle) and an unstable improper node (dots)) correspond to those in the polar coordinate case \cref{fig:pp_a1k0} given by the dashed dotted lines. The thin solid curve denotes the separation of the domain by $\xi=0$. Thick lines starting at singular points are selected solutions.
	      }
	\label{fig:pp_a2k0}
  \end{figure}

    The discussion for polar and spherical coordinates and a constant initial density distribution requires more effort. We plot the phase diagrams \cref{fig:pp_a1k0,fig:pp_a2k0} together with the singular lines $u\left(\xi\right)=\left\{0\wedge\xi\pm 1\right\}$ and the singular points at $\mathcal{D}_1=\left(0,0\right)$ for both coordinates. These singular points are both saddle points with their stable solutions receding along the eigenvector $\boldsymbol{x}_2=\left(1,0\right)^\mathrm{T}$ and their unstable solutions leaving along $\boldsymbol{x}_1=\left(0,1\right)^\mathrm{T}$.
    The points $\mathcal{D}_2=\left(1,0\right)$ show different behavior. For $\alpha=1$ we find an unstable improper node as a source with its first eigenvector $\boldsymbol{z}_1=\left(1,0\right)^{\mathrm{T}}$ giving the main direction for the solutions to swell. All other solutions leave this point in direction of $\boldsymbol{z}_2=\left(2,1\right)^{\mathrm{T}}$. The correct ones are hence given with the one connecting the singular points and else satisfying the outer BC \cref{eq:ubnd}. The discontinuity must be located between the point $\mathcal{D}_2$ and the intersection between the solution and the singular line.

    The same is true in the case of $\alpha=2$. However, we find an unstable degenerate improper node, meaning there exists just one eigenvalue with multiplicity two and just one eigenvector $\boldsymbol{z}_1=\left(1,0\right)^{\mathrm{T}}$. The solution starting at the point $\mathcal{D}_2$ and entering the origin is stable. The BC again yields the correct upstream solution.
    
\begin{figure}
  \centerline{\includegraphics[angle=0]{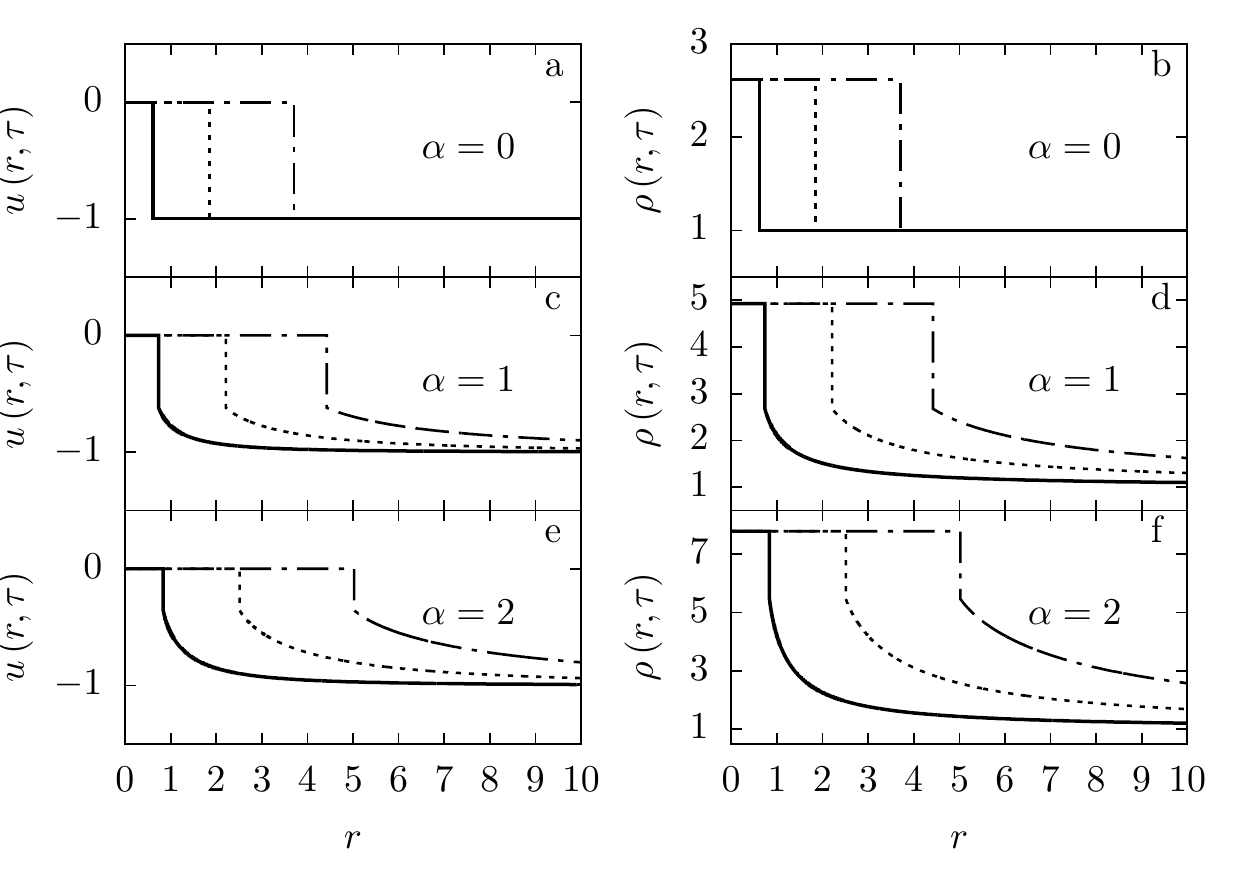}}
      \caption{Summarized solutions for the velocity (a,c,e) and the density (b,d,f) in all three coordinates ($\alpha\in\left\{0,1,2\right\}$) and $\kappa=0$ at different times $\tau$. The solid line is $\tau=1$, the dotted line is $\tau=3$ and the dash-dotted line is $\tau=6$.
	      }
	\label{fig:solk0comb}
    \centerline{\includegraphics{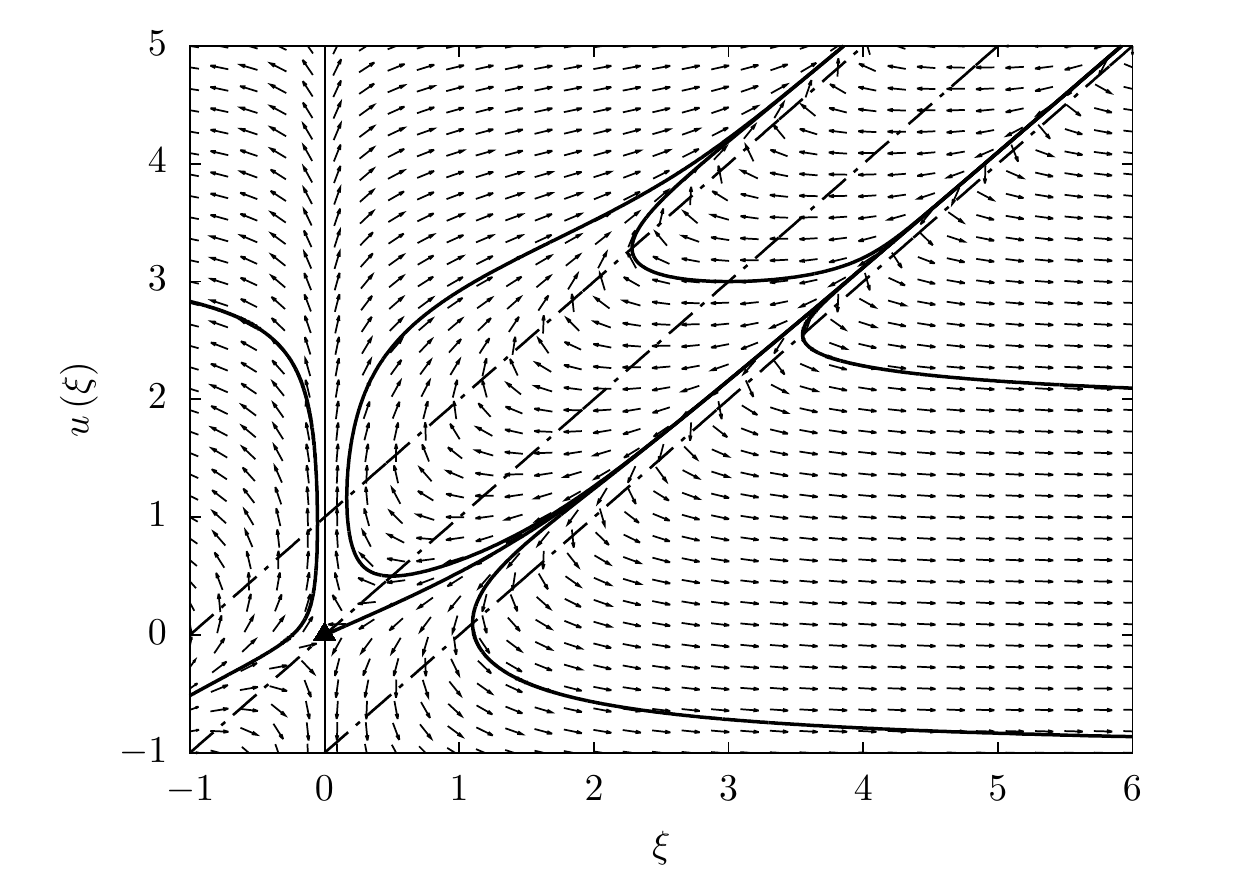}}
    \caption{Phase diagram for polar coordinates with ($\alpha=1, \kappa=1$). There is just one saddle point (triangle) and the thick line receding the origin is the real solution, whilst the others are some possible ones. The singular lines again are dashed dotted, the vertical line departs the phase plane ($\xi=0$).
		}
    \label{fig:pp_a1k1}
  \end{figure}

    All the solutions concerning different coordinates $\alpha\in\left\{0,1,2\right\}$ together with the constant initial density distribution are summarized in \cref{fig:solk0comb}a-f. The upstream velocity remains zero up to the discontinuity caused by the nature of the singular points. Hence, the density is forced to behave similarly because the right hand side of \Cref{eq:odes} vanishes for zero velocities.


 \subsubsection{Polar coordinates and \texorpdfstring{$\kappa=1$}{}}\label{sec_a1k1}
   We show at the end of \cref{anasolIP} that there are no possible solutions for $\alpha=0,~\kappa\geq1$, because the RHCs are never satisfied for that case. Hence, we start with the discussion for polar coordinates.
 
  The singular lines are given by $u\left(\xi\right)=\left\{\xi \wedge \xi\pm1\right\}$ and plotted in \cref{fig:pp_a1k1}. However, there exists just one singular point $\mathcal{D}_1\left(0,0\right)$, a saddle point gathering the stable solution along the direction of $\boldsymbol{x}_2=\left(2,1\right)^\mathrm{T}$. All three singular lines are parallel to each other. This is expressed in the diverging behavior of the singular point $\mathcal{D}_2$ \cref{eq:singpnt} which therefore does not exist. The correct solution describing the IPs upstream part leaves the stable solution branch at some point, so that the BCs are satisfied. The jump location is just restricted by the area behind the intersection of the upstream solution with the singular line. The solutions are plotted in \cref{fig:figure7}a,b. The velocity behind the origin rises rapidly up to the RHC and is caused by the geometry.

  \subsubsection{Spherical coordinates and \texorpdfstring{$\kappa=1$}{}}\label{a2k1}
\begin{figure}
    \centerline{\includegraphics{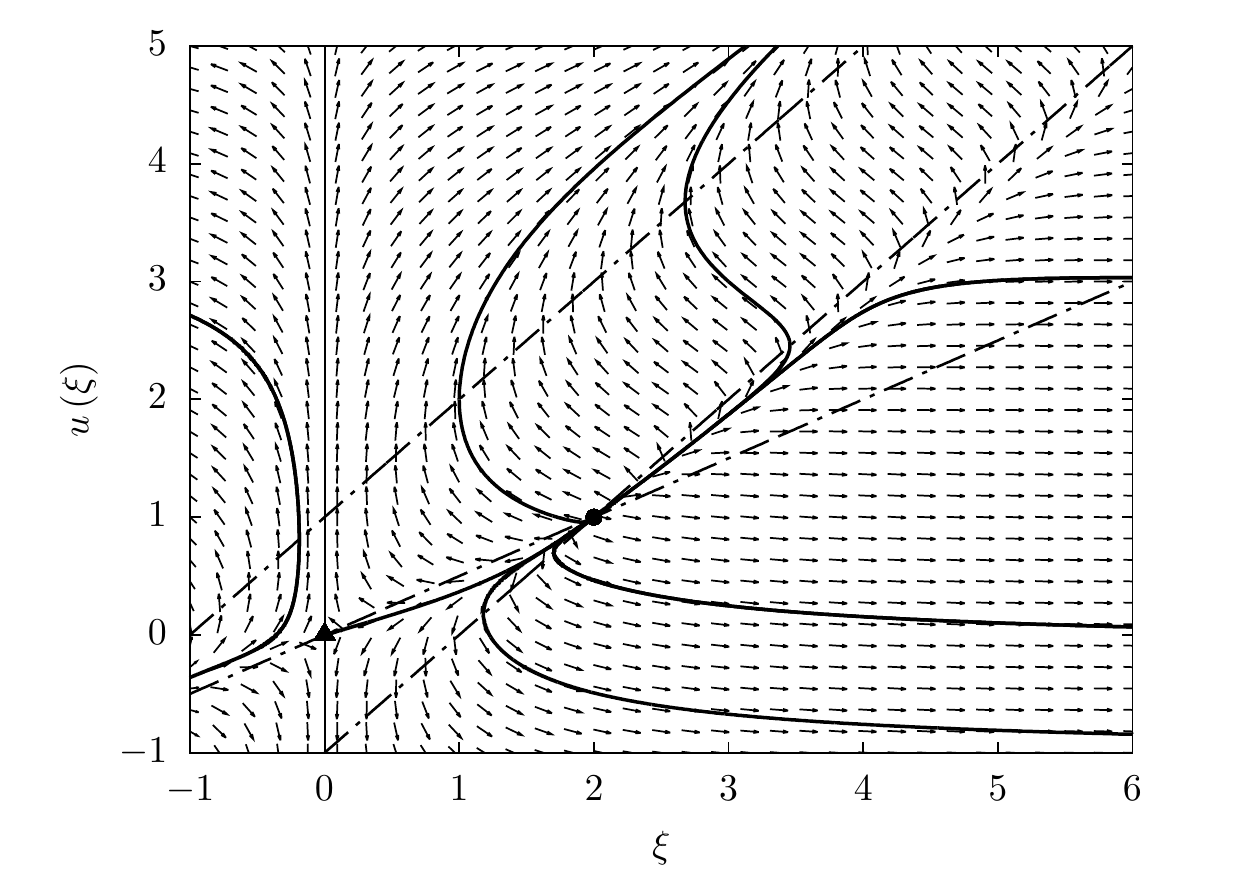}}
	\caption{Phase diagram for spherical coordinates $\alpha=2$ and $\kappa=1$. We find a saddle point (triangle) and an unstable improper node (dot) where the solution (solid line) connecting these points is a stable solution. The others are possible solutions, and the dash dotted are the singular lines.}
	  \label{fig:pp_a2k1}
   \end{figure}

\begin{figure}
     \centerline{\includegraphics{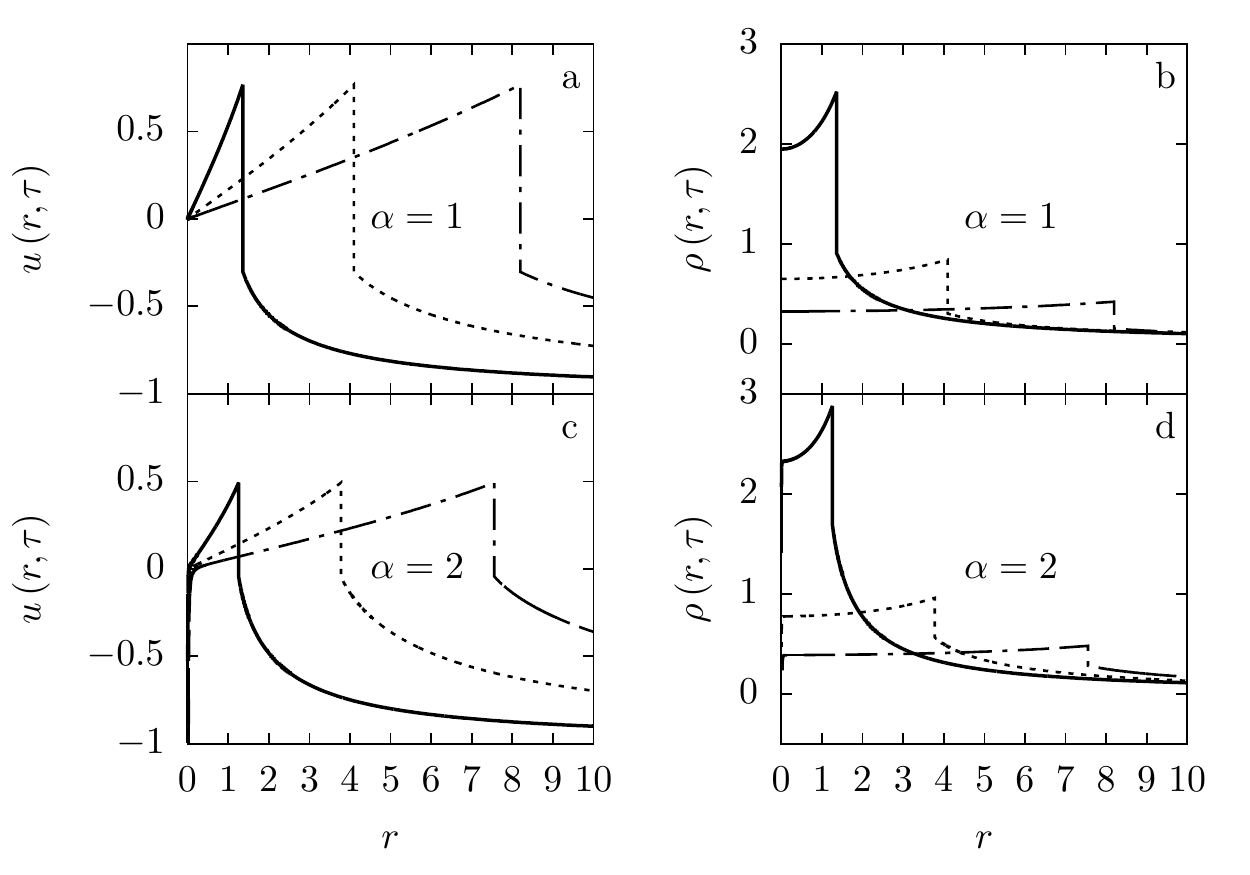}}
      \caption{Summarized solutions for the velocity (a,c) and the density (b,d) in different coordinates ($\alpha=1,2$) and $\kappa=1$ at different times $\tau=1,3,6$ (solid, dashed and dashed dotted lines, respectively). The behavior at the origin is caused by numerical uncertainties by computing the RHC.
	      }
     \label{fig:figure7}
    \end{figure}

   In this case, the singular lines in \cref{fig:pp_a2k1} are given by $u\left(\xi\right)=\left\{\xi/2 \wedge \xi\pm1\right\}$. Solutions receding from the saddle point $\mathcal{D}_1=\left(0,0\right)$ are stable along the vector $\boldsymbol{x}_2=\left(3,1\right)^{\mathrm{T}}$ describing the downstream part. This solution originates from the unstable improper node $\mathcal{D}_2=\left(2,1\right)$ leaving in the main direction of $\boldsymbol{z}_2=\left(-1+\sqrt{5},1\right)^{\mathrm{T}}$. The second direction is given by $\boldsymbol{z}_1=\left(-1-\sqrt{5},1\right)^{\mathrm{T}}$. The upstream part again must reach the BC defining a intersection with the singular line. Again, the location of the discontinuity is restricted between this intersection and the improper node. The plots of the solutions are shown in \cref{fig:figure7}c,d for different times $\tau$.
  
  \subsubsection{General restrictions for \texorpdfstring{$\kappa$}{}}\label{secgenrestrict}
    We argue at the end of \cref{anasolIP} that for $\kappa\geq1$ the RHCs yield no physically appropriate solutions for this regime. Moreover, we showed an easy method for solving \Cref{eq:odev,eq:odes} without any restriction to $\kappa$. But there is at least one we need to mention. Regarding \Cref{eq:odev} again, it is easy to spot and to prove that $u\left(\xi\right)=\xi$ (see end of \cref{anasolIP}) is a singular solution. We can isolate the system parameters $\alpha, \kappa$ due to this solution and find $\kappa=\alpha+1$. This relation restricts the parameter $\kappa$ in dependence of the geometry $\alpha$ inasmuch as every singular solution violates the RHCs \cref{eq:RHCV}. Hence, we are able to give the limit
    \begin{equation}
     \kappa<\alpha+1.
    \end{equation}
    for every possible solution. On a physical argumentation however, a finite initial mass is required to state a proper initial condition $\rho_0$ (\Cref{eq:IC_out}). By solving the mass integral $\left(M=c\int_0^r\rho_0r'^{\alpha}\mathrm{d}r'\right)$ for \Cref{eq:IC_out} and for the critical value $\kappa=\alpha+1-\epsilon$, $\epsilon\geq0$ we find
    \begin{equation}
      M=cS_\infty\int_0^r\frac{\mathrm{d}r'}{r'^{1-\epsilon}}=\frac{cS_\infty}{\epsilon}r^{\epsilon}\label{eq:mass}
    \end{equation}
    where $c$ is a constant depending on the geometry $\alpha$. The constant $\epsilon$ is introduced to summarize the behavior of the solutions. Whilst the integral diverges for $\epsilon=0$ and gives negative masses for $\epsilon <0$ valid solutions are given for $\epsilon>0$. Hence, together with $\kappa\in\mathbb{R}_0^+$ we conclude that
    \begin{equation}
     \kappa_\mathrm{max}<\alpha+1\label{eq:strict_limit}
    \end{equation}
    is a strict upper limit for the solveability of the system \cite{Solinger1975,Lerche1976}. These general findings remain also valid for the EP we discuss in the next section.

  \subsection{Isothermal blast waves (BW)}\label{secBW}
    In analogy to the well known solutions of point explosions by \cite{Sedov1946} and \cite{Taylor1950} the isothermal counterpart is derived by \cite{Korobeinikov1956}. He used the same assumption of a vanishing temperature gradient ahead of the shock as \cite{Sedov1946}, meaning $0\approx\partial_r T_{\mathrm{d}} \ll \partial_r T_{\mathrm{u}}$. But this assumption means that there exists a temperature jump $0=T_\mathrm{d}\neq T_\mathrm{u}$ in contrast to a real isothermal assumption where very efficient cooling causes the same temperature throughout the system $\left(T_\mathrm{d}=T_\mathrm{u}\right)$. Hence, this results in a mixture of adiabaticity and isothermality which is not genuinely isothermal. As a result there exists an adiabatic flow through the discontinuity and the principle of energy conservation therefore holds, even in this {\it{isothermal}} case. The similarity variable yields $\eta=\left(r^5\rho t^{-2}E^{-1}\right)^{1/5}$ with $E$ the initial energy, $t$ the time, $r$ the radial distance and $\rho$ the initial density. This enables the immediate release of a specified amount of energy at even very small spatial extent. Hence, there are well defined solutions for point-like explosions e.g. \cite{Korobeinikov1956, Solinger1975, Lerche1976} with respect to the initial parameters energy and density. In this chapter we want to develop a concept of genuinely isothermal BWs without the aforementioned mixed adiabatic and isothermal assumption. 

    We already deduced genuinely isothermal equations from the system \cref{eq:odev,eq:odes}. The only difference between the IPs treated earlier and the BWs is found in the initial conditions. 
\subsubsection{Spherical isothermal BW}
\begin{figure}
    \centerline{\includegraphics{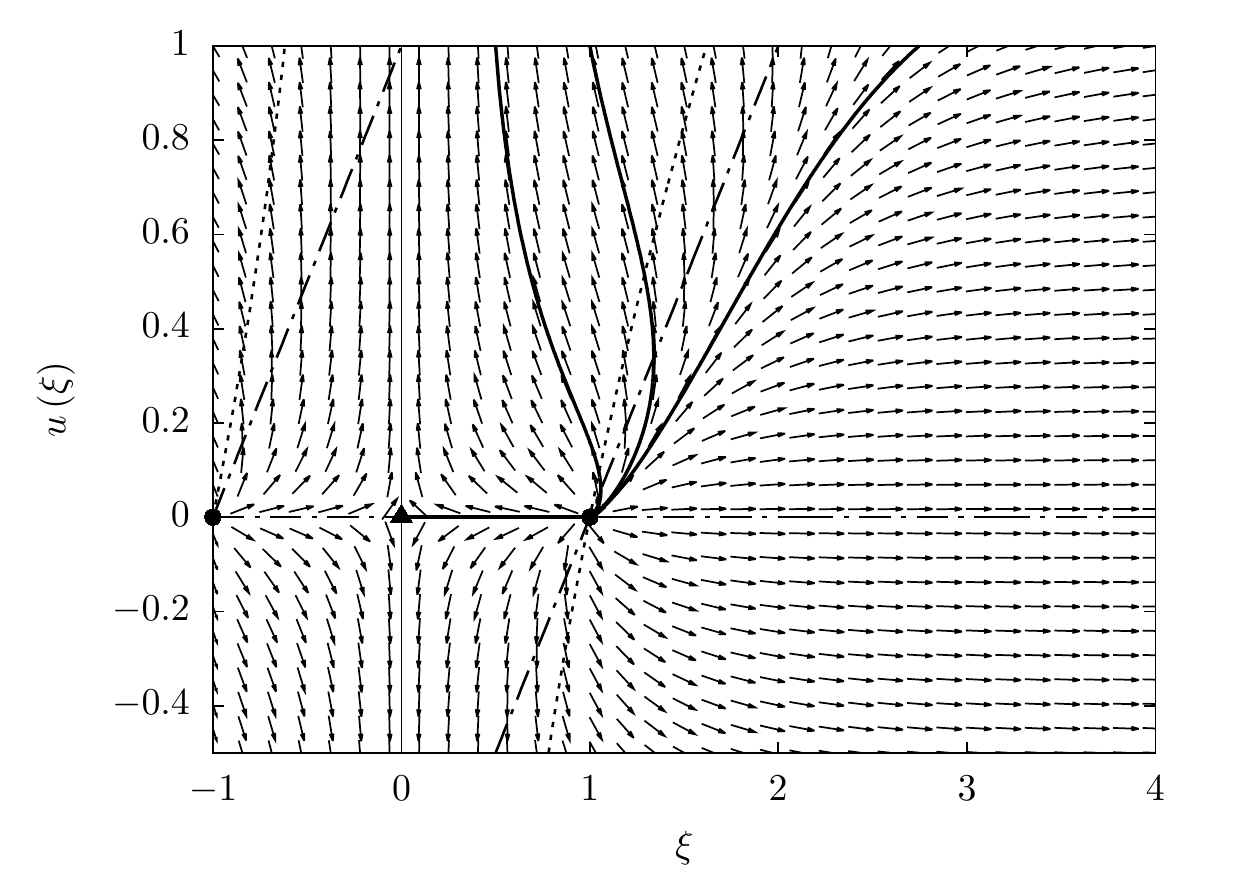}}
	\caption{Phase diagram for the EP ($\alpha=2, \kappa=0$). We spot a saddle point (triangle) and an unstable degenerate improper node (dots) where the solution (solid) connecting these points is stable. The dotted line gives the values $u_\mathrm{d}\left(\xi_0\right)$ \cref{eq:RHCBW} as a function of the shock position. The possible solutions do never satisfy the RHC without losing uniqueness. Hence, genuinely isothermal BW solutions are only valid for $\kappa>0$.
		}
	  \label{fig:pp_blast_k0}
    \end{figure}

    In the following, we discuss solutions in spherical coordinates ($\alpha=2$). Although the values for the RHCs \cref{eq:RHCV,eq:RHCS} are computed with the same routine as described earlier, there is a little difference due to the BC $\left(u_\infty=0\right)$, and hence the simple relations depend only on the shock position
    \begin{equation}
    	u_\mathrm{d}=\xi_0-\xi_0^{-1},~~S_\mathrm{d}=S_\mathrm{u}\xi_0^{-2}.\label{eq:RHCBW}
    \end{equation}	  
    In \cref{fig:pp_blast_k0} we plot the phase diagram for a constant initial density distribution ($\kappa=0$). Every possible solution leaving $\mathcal{D}_2=\left(1,0\right)$ either reaches its maximum $\xi$-value at the intersection with the singular line $u\left(\xi\right)=\xi-1$ and loses uniqueness afterwards, or never intersects at all. Further, the values from \Cref{eq:RHCBW} plotted as a function of the shock position $u_\mathrm{d}\left(\xi_0\right)$ (dotted line in \cref{fig:pp_blast_k0}) lies always above the singular line, meaning that no solution exists referring to $\kappa=0$ satisfying the RHCs without losing uniqueness. 

\begin{figure}
    \centerline{\includegraphics{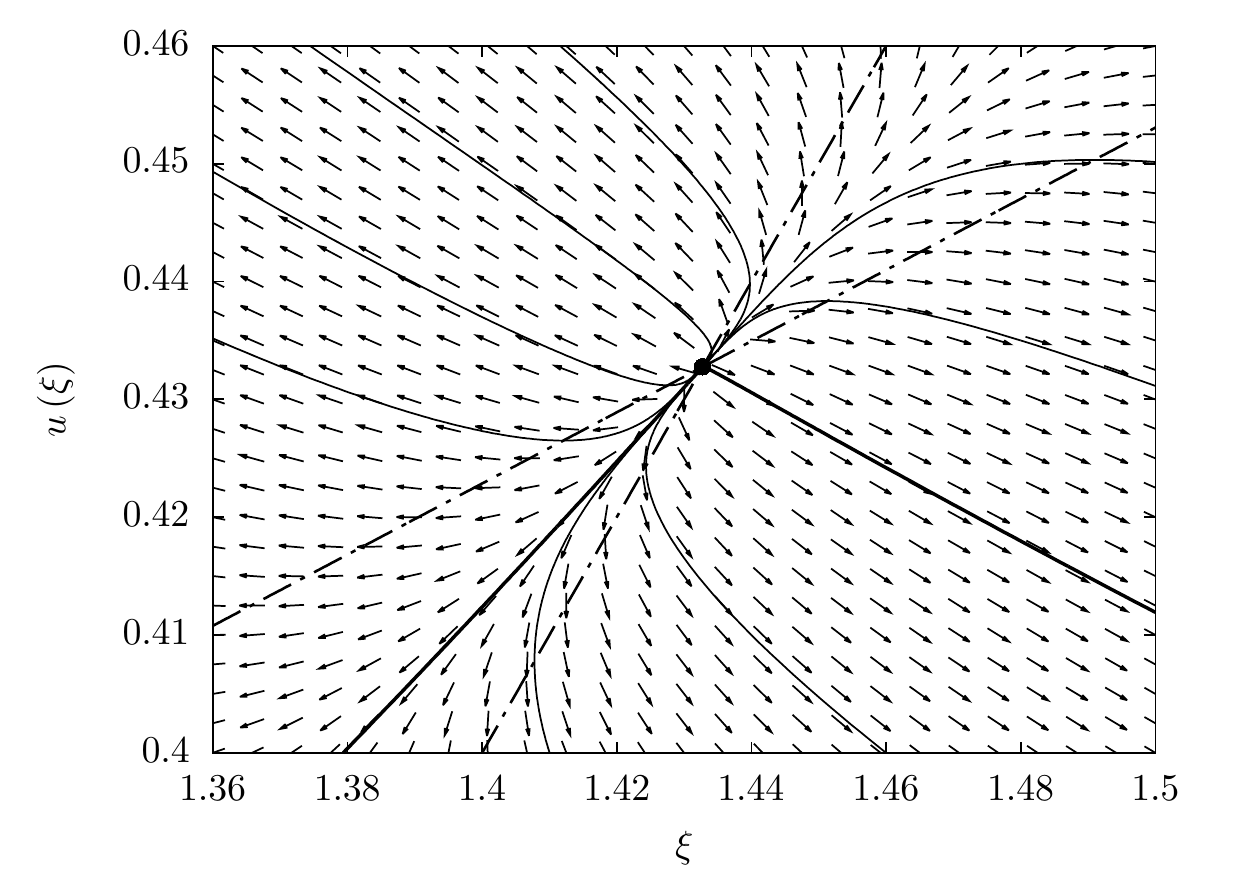}}
	\caption{A removable discontinuity satisfying the outer BC is found for $\kappa=0.60409$. This value denotes a lower restriction for the BW solutions.
		}
	  \label{fig:pp_kink}
    \centerline{\includegraphics{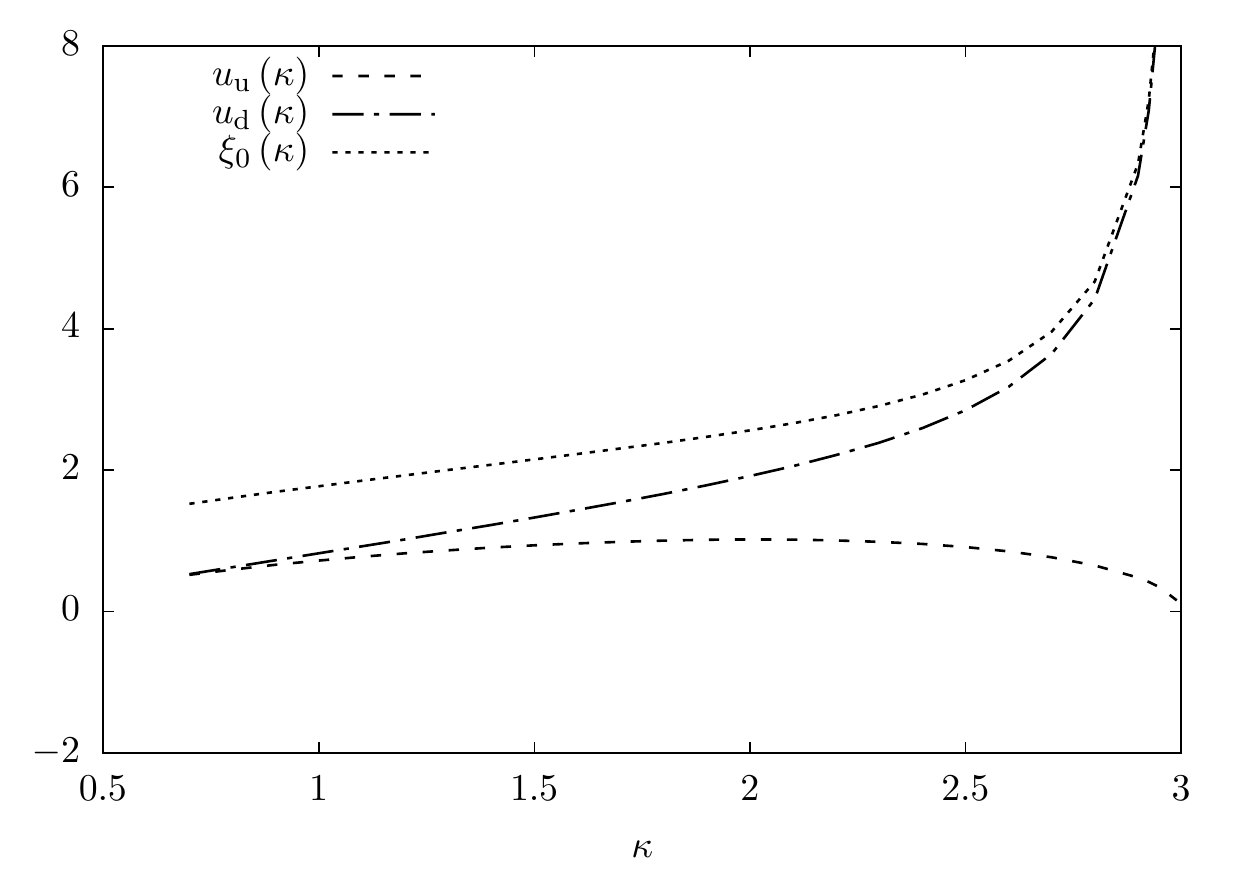}}
	\caption{The quantities $u_\mathrm{u}, u_\mathrm{d}$ and $\xi_0$ in dependence of $\kappa$. We again find the diverging behavior for $\kappa\rightarrow3$ in accordance to the results from \cref{secgenrestrict}. The vanishing jump is due to a removeable discontinuity in the solutions for $\kappa=0.60409$.
		}
	  \label{fig:blast_kdep}
    \end{figure}

\begin{figure}
    \centerline{\includegraphics{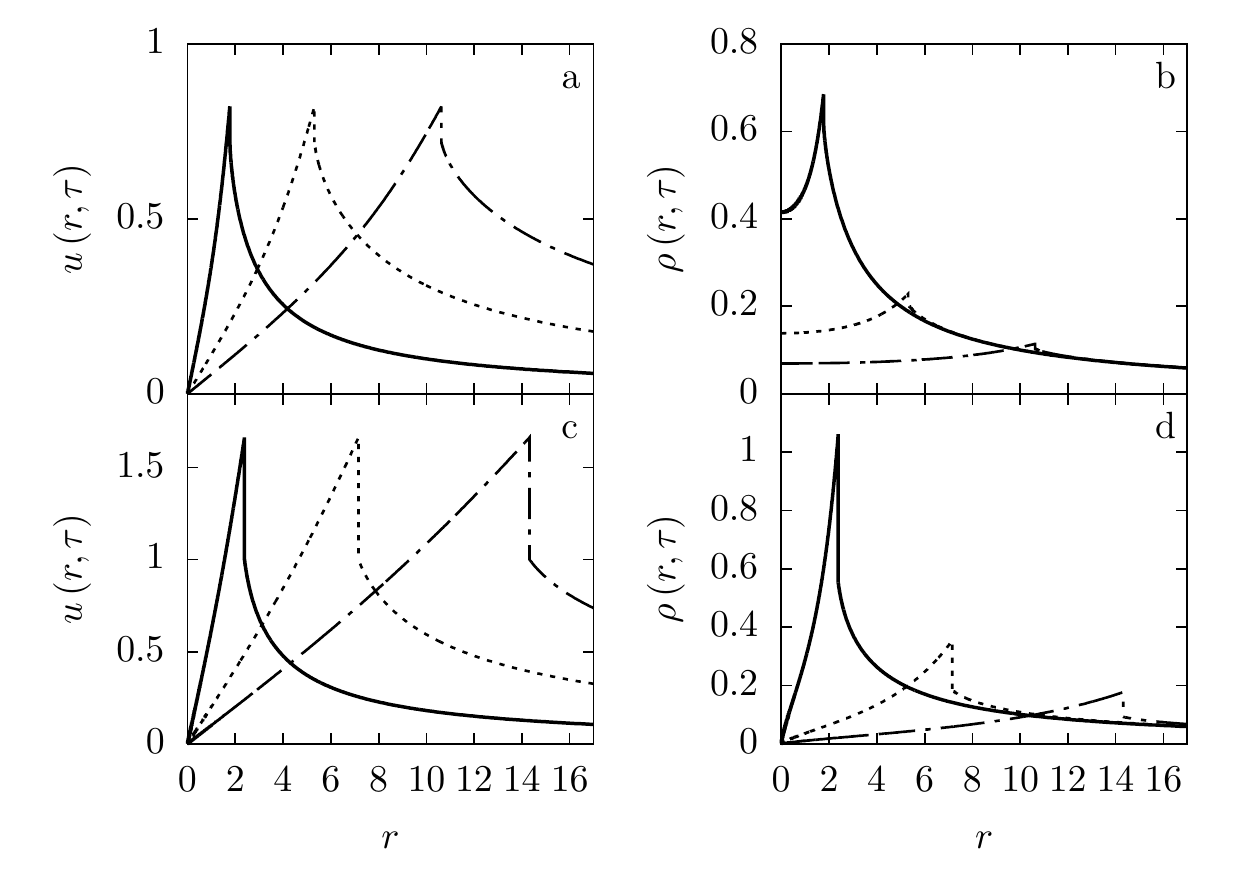}}
	\caption{BW solutions for $\alpha=2$ at different times $\tau=1$ solid, $\tau=3$ dashed and $\tau=6$ dashed dotted for $\kappa=1$ a,b and $\kappa=9/5$ c,d. The latter case yields a comparison due to the shock strength with \cite{Korobeinikov1956}.
		}
	  \label{fig:sol_blast}
    \end{figure}

    But there are further restrictions given. On the one hand, the strict upper limit \cref{eq:strict_limit} holds, meaning that the values are diverging for $\kappa\rightarrow3$. On the other hand, we derive a lower limit for $\kappa$ by searching for a removeable discontinuity (see \cref{fig:pp_kink}). This solution exists as also seen in \cref{fig:blast_kdep} meaning $\mathcal{U}_\mathrm{d}-\mathcal{U}_\mathrm{u}=0$ and hence showing just a kink. The linearization \cref{eq:spanalysis} around $\mathcal{D}_2$ yields a starting point for the kinked solution. $\kappa$ is now varied numerically until the outer BC $u_{\infty}=0$ is satisfied. Hence, genuinely isothermal BW solutions in spherical coordinates only exist in the range
    \begin{equation}
     0.60409\leq\kappa<3. \label{eq:restrkappa}
    \end{equation}
    \begin{table}
    \caption{Numerical values for the BW RHCs in spherical ($\alpha=2$) and plane coordinates ($\alpha=0$) with different numbers for $\kappa$.}
      \label{tab:numerRHCblast}
\centering
    \def~{\hphantom{0}}
      \begin{tabular}{|c|c|c|c|c|c|c|}\hline
	  $\kappa$ & $\alpha$ & $\xi_0$ & $u_{\mathrm{u}}$ & $u_{\mathrm{d}}$ & $S_{\mathrm{u}}$ & $S_{\mathrm{d}}$\\ \hline
	  1/3 & 0 & 1.5844 & 0.3100 & 0.7998 & 1.1743 & 1.9072 \\ 
	  1/2 &   & 1.8694 & 0.3590 & 1.2073 & 1.1835 & 2.6998 \\ 
	  2/3 &   & 2.2696 & 0.3605 & 1.7458 & 1.1587 & 4.2229 \\ 
	    1 & 2 & 1.7716 & 0.7192 & 0.8213 & 1.0929 & 1.2103	\\ 
	  9/5 & 2 & 2.3858 & 1.0023 & 1.6629 & 1.3182 & 2.5225 \\ \hline 
      \end{tabular}
    \end{table}
    In the following we discuss the value $\kappa=1$ and use the phase diagram \cref{fig:pp_a2k1}. The singular points and their properties are already known from \cref{a2k1} yielding a saddle point at $\mathcal{D}_1\left(0,0\right)$ and an unstable improper node at $\mathcal{D}_2\left(2,1\right)$. The permitted solutions are those leaving $\mathcal{D}_2$ and entering either $\mathcal{D}_1$ or satisfying the corresponding outer BC \Cref{eq:ubnd}. Together with the numerical values of the RHC \cref{eq:RHCV,eq:RHCS} (see \cref{tab:numerRHCblast}), we have sufficient information to give well-defined solutions to the isothermal BWs presented in \cref{fig:sol_blast}a,b. \cref{fig:sol_blast}c,d demonstrate the impact of $\kappa=9/5$ on the velocity and the density.
    
    The solutions derived by \cite{Korobeinikov1956} assume constant initial energy and density. Thus, a point-like source or a delta peak as initial condition is set and the explosion is triggered by the energy which is instantaneously released. \cite{Lerche1976,Sedov1959,Solinger1975} introduced an ambient initial density depending on a power law of the radial distance $\left(\rho\propto r^{-\omega}\right)$ artificially. This is very similar to our results for the initial density distribution, whereas \cref{eq:IC_out} results naturally from the stretching transformation \cref{eq:densdist}. Nevertheless, it is argued in \cite{Solinger1975} that this parameter has to be zero for physically acceptable solutions, whilst we determined $\kappa>0$ necessarily for a genuinely isothermal BW. In \cite{Sedov1959} a variable initial density is investigated. It is shown, that the exponent of the density $\omega<3$ must hold for a finite mass in the system, similar to our results in \cref{secgenrestrict}. Further, we adjust the initial mass instead of the initial energy and trigger the explosion by the mass distribution due to $\kappa$. The shock velocity, derived by the time derivative of the shock position $v_\mathrm{s}=\xi_0\left(\kappa\right)$ \cref{eq:simvar} and \cref{fig:blast_kdep} may explain this behavior. The steeper the density gradient (meaning increasing values for $\kappa$) the higher the shock velocity and the stronger the shock appears. This finding is also shown in \cite[p. 260]{Sedov1959}.
    The solution for the velocity in \cite{Korobeinikov1956} equals zero between the origin (due to the BCs) and the first singular point. This behavior is somehow astonishing because there is no obvious physical reason for the gas to stay at rest until it reaches the first singular point. Moreover, we discard the solutions for $\kappa=0$ in \cref{fig:pp_blast_k0} (due to relation \cref{eq:restrkappa}) showing similar behavior between the origin and the singular point $\mathcal{D}_2$. This does not say that this behavior is per se wrong. It rather shows another difference between the previously made mixed adiabatic and isothermal and genuinely isothermal solutions.

  \section{Summary}\label{sec:summary}
    We present solutions for the 1D isothermal IP and EP assuming slab-, polar- and spherical-symmetry, respectively. The derivation of the ODE leads to both solutions by just changing the initial conditions. Analytical expressions for the cartesian IP and EP are derived which we use to verify the correctness of the numerical integration routines for the non-analytical solutions yielding very good agreement. We give a mathematical motivation for the similarity variable and hence reduce the possible solution branches to those beeing proportional to some power law depending on radius for the initial density distribution.

    A genuinely isothermal ansatz is used, meaning the same temperature throughout the system. This is a basic difference to the isothermal solutions by \cite{Korobeinikov1956}, where the temperatures in the up- and downstream regions are different, causing an adiabatic flow through the discontinuity. The mass in our system is tuned instead of a constant initial energy gain, whereby the shock strength is triggered by the exponent $\kappa$ of the radial distance of the initial density distribution. Although the parameter $\kappa>0$ is a necessary requirement for genuinely isothermal BWs in spherical coordinates we further found a restriction in the range of $0.60409\leq\kappa<3$.

    Our solutions should be checked in an astrophysical context like supernovae- or collapse events. Due to the isothermality one can imagine two different regimes. On the one hand, we may describe an IP/ EP event happening in an optical thin regime, where the timescales for radiative cooling are much smaller than those for shock heating processes. On the other hand, IP/ EP may be embedded in a hot environment (e.g., a H\scriptsize{II} \normalsize region) \cite{Shu2002}, where the shock heating does not change the temperature significantly. The genuinely isothermal solutions are appropriate, even if we do not have any external forces like gravitation (as used in other works e.g.,) \cite{Shu1992, Shu2002} as triggering forces.

    Another reasonable application lies in the usage as a test case for numerical hydrodynamics codes. In particular, flux transport in the radial direction and the mass conservation are crucial characteristics of these codes and play an important role in our exact solutions \cite{Illenseer2009}.

  \appendix
  
  \section{Analytical solutions for the IP in 1D planar flow}\label{anasolIP}
  We set $\alpha=0$ for cartesian coordinates and insert the definition $w\doteq \xi-u$ into \cref{eq:odev} leading to the separable equation
  \begin{equation}
    \frac{\mathrm{d}w}{\mathrm{d}\xi}\left(1+\frac{\kappa}{\left(1-\kappa\right)-w^2}\right)=1.\label{eq:odeanalytu}
  \end{equation}
  The fractal in the brackets changes its behavior if $\kappa=1$. Hence, we investigate the solutions for $\kappa<1$ and discuss the case for $\kappa\geq1$ at the end of this section. The general solution of \cref{eq:odeanalytu} for $\kappa<1$ is given by
  \begin{equation}
    w+\frac{1}{\nu}\ln\left(\pm\frac{\sigma+w}{\sigma-w}\right)=\xi+c_\pm\label{eq:genalayt}
  \end{equation}
  with $\nu=\frac{2\sqrt{1-\kappa}}{\kappa}$ and $\sigma=\sqrt{1-\kappa}$ and $c_\pm$ the constants of integration. The positive sign applies for $\left|w\right|<\sigma$ and the negative sign otherwise. Thus, the solution splits into two branches which one can relate to the upstream and downstream solutions (see \cref{fig:Figtube})
  \begin{align*}
    u\rightarrow 0 \mbox{\ for\ } \xi\rightarrow0 &\Rightarrow\left|w\right|\rightarrow 0<\sqrt{1-\kappa} &\mbox{(downstream)}\\
    u \rightarrow -1 \mbox{\ for\ } \xi\rightarrow\infty &\Rightarrow\left|w\right|\rightarrow\infty>\sqrt{1-\kappa}. &\mbox{(upstream)}
  \end{align*}
  If we apply the BC (\Cref{eq:appr_u,eq:ubnd}) the constants of integration become $c_{+}=1$ and $c_{-}=e^\nu$. Hence, we obtain implicit expressions for the up- and downstream solutions
  \begin{subequations}
  \begin{align}
    \xi\left(u\right)&=u-\sigma\frac{1-e^{\nu u}}{1+e^{\nu u}}~~,~~~~~~~~\xi<\xi_0 \label{eq:analytd}\\
    \xi\left(u\right)&=u-\sigma\frac{1+e^{\nu \left(1+u\right)}}{1-e^{\nu \left(1+u\right)}}~~,~~~\xi>\xi_0\label{eq:analytu}
  \end{align}
  \end{subequations}
  where $\xi_0$ is the position of the shock. \Cref{eq:analytd,eq:analytu} together with the RHC \cref{eq:RHCV} yield three equations for the three unknowns ($u_\mathrm{d}, u_\mathrm{u}, \xi_0$). To solve this system we define
  \begin{equation}
   \psi\doteq\frac{1-e^{\nu\left(1+u_\mathrm{u}\right)}}{1+e^{\nu\left(1+u_\mathrm{u}\right)}}~~\Leftrightarrow~~ u_\mathrm{u}=\frac{1}{\nu}\ln\left(\frac{1-\psi}{1+\psi}\right)-1.\label{eq:psi}
  \end{equation}
  with the upstream velocity $u_\mathrm{u}=\lim_{\xi\rightarrow\xi_0^+} u^{\left(\xi>\xi_0\right)}$ being valid at the jump position $\xi_0$. Insertion of this relation into \Cref{eq:analytu} gives the jump position
  \begin{equation}
   \xi_0=u_\mathrm{u}-\frac{\sigma}{\psi}=u_\mathrm{d}-\frac{\psi}{\sigma}.\label{eq:xiu}
  \end{equation}
  Thereby we use the RHC \cref{eq:RHCV} to obtain the second relation. Furthermore, we express the downstream velocity $u_\mathrm{d}=\lim_{\xi\rightarrow\xi_0^-} u^{\left(\xi<\xi_0\right)}$ at the shock in terms of $\psi$ using \Cref{eq:analytd,eq:xiu} and the RHC \cref{eq:RHCV} yielding
  \begin{equation}
    u_\mathrm{d}=\frac{1}{\nu}\ln\left(\frac{\sigma^2-\psi}{\sigma^2+\psi}\right).\label{eq:psid}
  \end{equation}
  With the help of \Cref{eq:psi,eq:psid} we eliminate $u_\mathrm{u}$ and $u_\mathrm{d}$ from \Cref{eq:xiu} and deduce the nonlinear conditional equation for $\psi$
  \begin{equation}
   f\left(\psi\right)=\frac{1}{\nu}\ln\left(\frac{\left(\sigma^2-\psi\right)\left(1+\psi\right)}{\left(\sigma^2+\psi\right)\left(1-\psi\right)}\right)-\frac{\psi}{\sigma}+\frac{\sigma}{\psi}+1=0.\label{eq:fpsi}
  \end{equation}
  Solving \cref{eq:fpsi} numerically for $\psi$ and inserting the result into \Cref{eq:psi,eq:xiu,eq:psid} yields the values $u_\mathrm{u},~u_\mathrm{d}$ and $\xi_0$ at the shock. We solved the conditional equation for different values of $\kappa=\left\{1/3, 1/2, 2/3\right\}$ and compute the relative error according to the numerical procedure described in \Cref{secRHC} which are summarized \Cref{tab:numerRHC_analy}. 
  
  The remaining differential equation \cref{eq:odes} is rewritten for functions $S\left(u\right)$ instead of functions $S\left(\xi\right)$ according to 
  \begin{equation}
    \frac{\mathrm{d}S}{\mathrm{d}u}=S\frac{1+u\left(\xi-u\right)}{\xi}\label{eq:odeanalyts}
  \end{equation}
  which can handle the implicit solutions given by \Cref{eq:analytd,eq:analytu}. This differential equation is also separable yielding the solutions
  \begin{subequations}
  \begin{align}
    S\left(u\right)&=C_\mathrm{+}e^{\mathcal{V}\left(u\right)},~~~\xi<\xi_0\label{eq:soldS}\\
    S\left(u\right)&=C_\mathrm{-}e^{\mathcal{W}\left(u\right)},~~~\xi>\xi_0\label{eq:soluS}
  \end{align}
  \end{subequations}
   for the up- and downstream domains, respectively. The functions in the exponents are given by the integrals
  \begin{subequations}
  \begin{align}
  	\mathcal{V}\left(u\right)&=\int_0^{u}\frac{\left(1+e^{\nu u'}\right)-u'\sigma\left(1-e^{\nu u'}\right)}{u'\left(1+e^{\nu u'}\right)-\sigma\left(1-e^{\nu u'}\right)}\mathrm{d}u'\label{eq:analytdensV}\\ 
  	\mathcal{W}\left(u\right)&=\int_{u}^{-1}\frac{\left(1-e^{\nu\left(1+u'\right)}\right)-\sigma u'\left(1+e^{\nu\left(1+u'\right)}\right)}{u'\left(1-e^{\nu\left(1+u'\right)}\right)-\sigma\left(1+e^{\nu\left(1+u'\right)}\right)}\mathrm{d}u'\label{eq:analyt_dens}
  \end{align}
  \end{subequations}
  which we cannot evaluate analytically. Nevertheless, we use the outer BCs \cref{eq:sbnd,eq:ubnd} and together with
  \begin{equation}
    \lim_{\xi\rightarrow\xi_\infty}S\left(\xi\right)=\lim_{u\rightarrow u_\infty} S\left(u\right)= S_\infty=1.
  \end{equation}
  and \Cref{eq:soluS} we can fix the constant of integration $C_-=1$. This enables us to evaluate the integral in \cref{eq:analyt_dens} for a given upstream velocity $u_\mathrm{u}$ at the shock and with help of \cref{eq:soluS} we get the corresponding upstream density $S_\mathrm{u}$. Application of the RHC \cref{eq:RHCS} yields again the downstream density $S_\mathrm{d}$ at the shock.
  \begin{figure}
    \centerline{\includegraphics{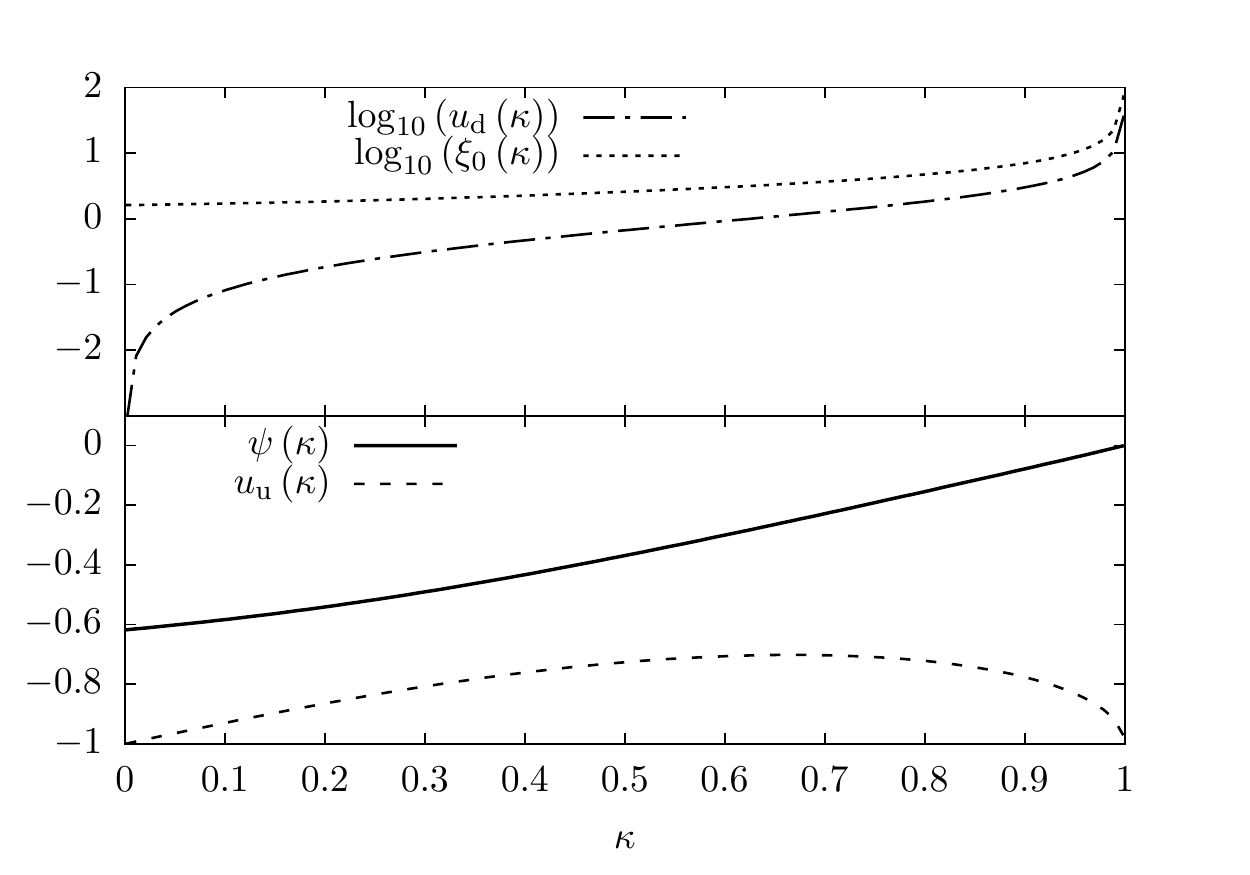}}
      \caption{The quantities $u_\mathrm{u}$, $u_\mathrm{d}$ and $\xi_0$ in dependence of the parameter $\kappa$. One easily spots their diverging behavior, especially for $u_\mathrm{d}\left(\kappa\rightarrow1\right)$ and $\xi\left(\kappa\rightarrow1\right)$.
	      }
	\label{fig:kdepen_analyt}
  \end{figure}
  \cref{fig:analyt} shows the velocity and the density solutions compared with the analytical expressions of the solutions calculated by solving the ODE \cref{eq:odev,eq:odes} numerically with the procedure we describe in \cref{kappa=0}. The analytical values are given in \Cref{tab:numerRHC_analy} (numerical values see \cref{tab:numerRHC}) and the relative error is summarized in \cref{tab:vergleichRHC}. Hence, we present two different methods yielding almost identical results and showing the quality of the numerical procedure.
  \begin{table}
  \caption{Analytical values for the RHC with different power law exponents of the initial density distribution $\kappa<1$ in slab coordinates ($\alpha=0$).}
    \label{tab:numerRHC_analy}
    \centering
      \def~{\hphantom{0}}
      \begin{tabular}{|c|c|c|c|c|c|}\hline
      $\kappa$ & $\xi_0$ & $u_{\mathrm{u}}$ & $u_{\mathrm{d}}$ & $S_{\mathrm{u}}$ & $S_{\mathrm{d}}$\\ \hline
      1/3 & 0.9389 & -0.7907 & 0.3607 & 0.8508 & 2.5449\\ 
      1/2 & 1.1905 & -0.7262 & 0.6687 & 0.8175 & 3.0031 \\ 
      2/3 & 1.5829 & -0.7017 & 1.1452 & 0.8004 & 4.1779 \\ \hline 
      \end{tabular}
  \end{table}
  \begin{figure}
  \centerline{\includegraphics{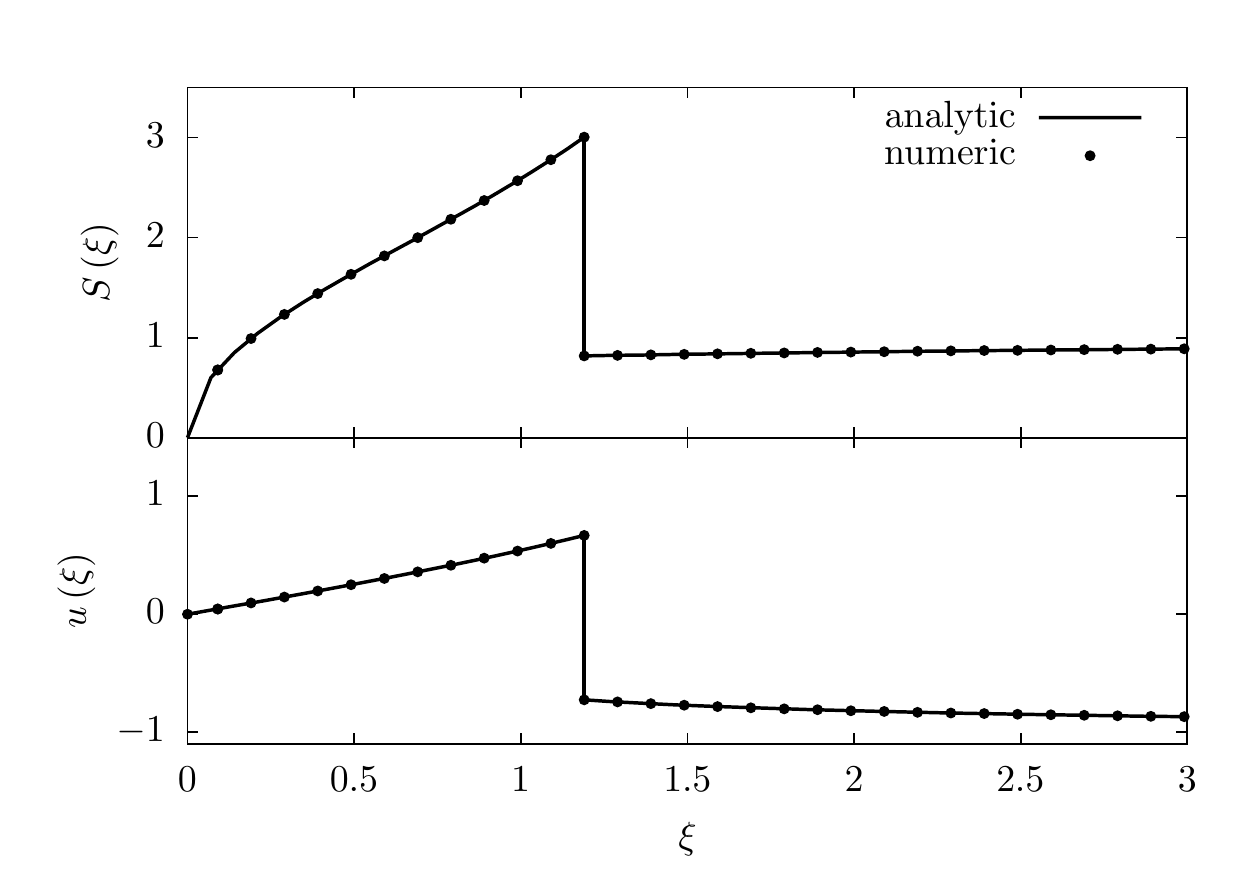}}
      \caption{Analytical solutions in comparison to the solutions derived by numerically solving the ODE \cref{eq:odev,eq:odes} described in \cref{kappa=0} using $\kappa=1/2$. The missing constant of integration for $\kappa=1/2$ is computed to $C_+=0.0972$. We find very good agreement between the solutions and also for the RHC given in \cref{tab:vergleichRHC}.
	      }
	\label{fig:analyt}
  \end{figure}
  
  In \cref{fig:kdepen_analyt} we plot \Cref{eq:psi,eq:xiu,eq:psid} as functions of the parameter $\kappa\in\left[0,1\right)$. The functions $u_\mathrm{d}\left(\kappa\right)$ and $\xi_0\left(\kappa\right)$ diverge for $\kappa\rightarrow1$ and hence exclude the value $\kappa=1$ from the ODEs \cref{eq:odeanalytu,eq:odeanalyts}.

  We are able to solve \Cref{eq:odeanalytu} for $\kappa=1$ yielding simple analytical expressions for the upstream as well as the downstream region. The downstream part however is only described reasonably by the singular solution \cite{Stepanow1976} $u_\mathrm{d}\left(\xi\right)=\xi$. But, a substitution of this solution into the RHC \cref{eq:RHCV,eq:RHCS} shows undefined behaviour, meaning there is no appropriate physical solution for $\kappa=1$ satisfying the RHCs.

  We also find implicit solutions of \Cref{eq:odeanalytu} for $\kappa>1$. We must, however, reject these solutions, because we can show that there are no physical appropriate solutions for any $\kappa>1$ in the domain of interest satisfying both the RHCs and the corresponding BCs.

\section{Analytical EP solutions for 1D planar flow} \label{anasolEP}
  Following the procedure from \cref{anasolIP} we find the general solutions in planar geometry
  \begin{equation}
    \xi\left(u\right)=u-\sigma\frac{1\mp e^{\nu u}}{1\pm e^{\nu u}}~~,~~~~~~~~\xi\lessgtr\xi_0\label{eq:analytdbw}
  \end{equation}    
  for the case $\kappa<1$ by application of the proper outer BC $u_\infty=0$. \Cref{eq:analytd,eq:analytdbw} with the upper sign are equal because the inner BC remains unchanged. Again using the RHC \cref{eq:RHCV} we derive a conditional equation for $\psi$
  \begin{equation}
    f\left(\psi\right)=\frac{1}{\nu}\ln\left(\frac{\left(1-\sigma^2\psi\right)\left(1+\psi\right)}{\left(1+\sigma^2\psi\right)\left(1-\psi\right)}\right)-\frac{1}{\sigma\psi}+\sigma\psi=0.\label{eq:fpsibw}
  \end{equation}
  and find the analytical expressions
  \begin{equation}
    u_\mathrm{d}=\frac{1}{\nu}\ln\left(\frac{1+\psi}{1-\psi}\right),~~u_\mathrm{u}=\frac{1}{\nu}\ln\left(\frac{1-\sigma^2\psi}{1+\sigma^2\psi}\right),~~\xi_0=u_\mathrm{d}-\sigma\psi=u_\mathrm{u}-\frac{1}{\sigma\psi}
  \end{equation}
  for the up- and downstream states at the discontinuity. 
  \begin{figure}
    \centerline{\includegraphics{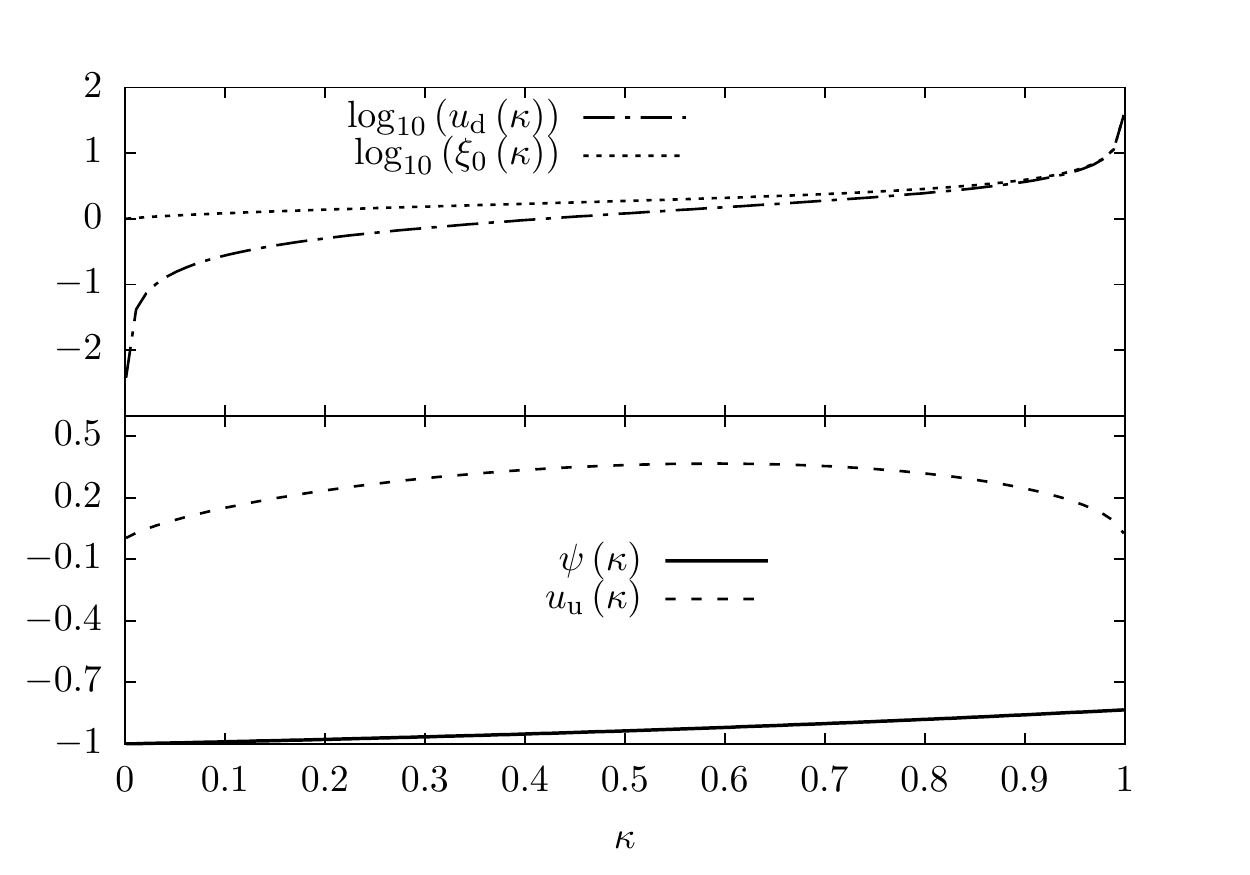}}
    \caption{The RHC in dependence of the parameter $\kappa$ in planar geometry ($\alpha=0$). The value $\kappa=1$ is also excluded in this regime because of the diverging behaviour $u_\mathrm{u,d}/\xi_0\left(\kappa\rightarrow1\right)\rightarrow\infty$.
    }
    \label{fig:BWanakdep}
\end{figure}
\begin{figure}
    \centerline{\includegraphics{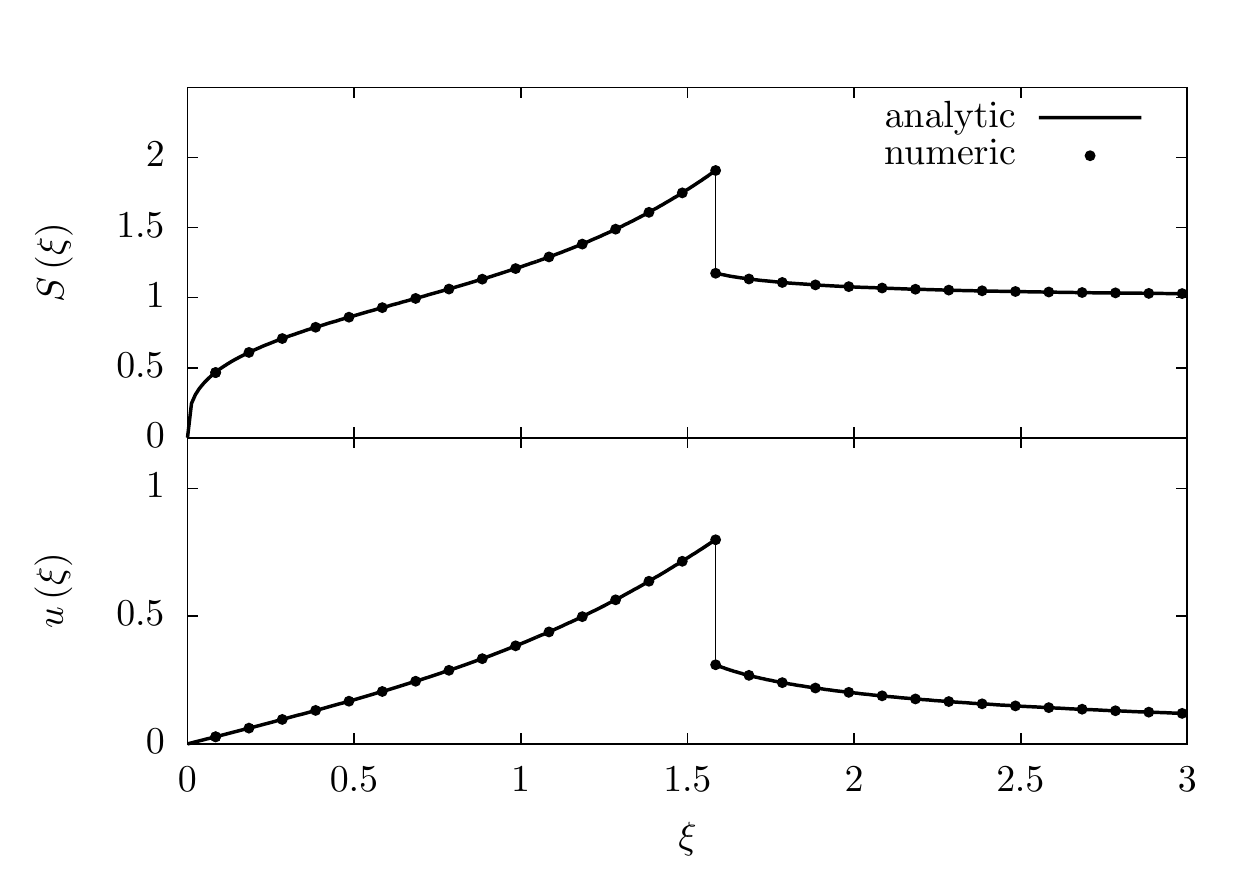}}
    \caption{Analytical solutions for the EP problem in planar geometry in comparison to the numerical results. The parameter $\kappa=1/3$ is shown and the integration constant $C_+=0.3388$ is derived. Again, there is very good agreement between the analytical and the numerical solutions.
    }
    \label{fig:BWanasol}
  \end{figure}
  \begin{table}
    \caption{Analytical values for the EP RHCs ($\alpha=0$) with different numbers for $\kappa$.}
    \label{tab:anaRHCblast}
    \centering
    \def~{\hphantom{0}}
      \begin{tabular}{|c|c|c|c|c|c|c|}\hline
	  $\kappa$ & $\alpha$ & $\xi_0$ & $u_{\mathrm{u}}$ & $u_{\mathrm{d}}$ & $S_{\mathrm{u}}$ & $S_{\mathrm{d}}$\\ \hline
	  1/3 & 0 & 1.5844 & 0.3100 & 0.7998 & 1.1743 & 1.9072 \\ 
	  1/2 & 0 & 1.8694 & 0.3590 & 1.2073 & 1.1835 & 2.6999 \\ 
	  2/3 & 0 & 2.2696 & 0.3605 & 1.7457 & 1.1587 & 4.2229 \\ \hline
      \end{tabular}
  \end{table}
  Analytical values for $\kappa=\left\{1/3,~1/2,~2/3\right\}$ are summarized in \cref{tab:anaRHCblast} (numerical values see \cref{tab:numerRHCblast}) and the relative error again shows very good agreement, see \cref{tab:vergleichRHC}. The solutions for the density yielding similar expressions as \Cref{eq:soldS,eq:soluS} where \Cref{eq:analyt_dens} changes to\footnote{The integral $\mathcal{V}\left(u\right)$ remains unchanged because of the same inner BC and is already given via \Cref{eq:analytdensV}.}
  \begin{equation}
    \mathcal{W}\left(u\right)=\int_{u}^{0}\frac{\left(1- e^{\nu u'}\right)-u'\sigma\left(1+ e^{\nu u'}\right)}{u'\left(1- e^{\nu u'}\right)-\sigma\left(1+ e^{\nu u'}\right)}\mathrm{d}u'.
  \end{equation}
  The density solutions and the corresponding constant of integration $C_+$ are again solved numerically. \Cref{fig:BWanakdep} shows the dependence of the RHC on $\kappa<1$. The value $\kappa=1$ is excluded because the RHC will again diverge for $\kappa\rightarrow1$. Solutions for $\kappa>1$ must be excluded, because they fail satisfying the RHCs (see \cref{anasolIP}). In \cref{fig:BWanasol} the analytical and numerical solutions for $\kappa=1/3$ are compared yielding very good agreement.

\bibliographystyle{siamplain}


	\bibliography{library_clean}


\end{document}